\documentclass[twocolumn,aps,prl,superscriptaddress,english]{revtex4-2}
\usepackage{amsmath,amsfonts, amssymb, amsthm}

\usepackage{bm}
\usepackage{mathrsfs}
\usepackage{graphicx}
\usepackage{verbatim}
\usepackage[colorlinks=true,citecolor=blue]{hyperref}
\usepackage[normalem]{ulem}
\usepackage{MnSymbol}
\usepackage{empheq}
\usepackage{tensor}
\usepackage{tikz}
\usetikzlibrary{calc}
\usepackage{mathtools}
\usepackage{adjustbox}
\usetikzlibrary {matrix}
\usetikzlibrary{positioning}
\usetikzlibrary{decorations.markings,arrows}
\usetikzlibrary { decorations.pathmorphing, decorations.pathreplacing, decorations.shapes}
\usetikzlibrary{shapes.geometric}
\usetikzlibrary{arrows.meta}
\usetikzlibrary {3d}
\usetikzlibrary {patterns,patterns.meta}
\usetikzlibrary {shadows}
\usepackage{natbib}
\usepackage{slashed}

\newcommand{\ee}{\mathrm{e}} 
\newcommand{\ii}{\operatorname{i}} 
\newcommand{\TT}{\mathcal{T}} 
\newcommand{\KK}{\mathcal{K}} 

\newcommand{\ZZ}{\mathbb{Z}} 
\newcommand{\HH}{\mathcal{H}} 

\newcommand{\Tr}{\mathrm{Tr}}

\newcommand{\dg}{\dagger}

\newcommand{\ket}[1]{\lvert#1\rangle}
\newcommand{\bra}[1]{\langle#1\rvert}

\newcommand{\lrangle}[1]{\langle#1\rangle}

\newcommand{\abs}[1]{\left\lvert#1\right\rvert}

\newcommand{\vket}[1]{\left| #1\right)}

\renewcommand{\tilde}{\widetilde}

\renewcommand{\geq}{\geqslant}
\hypersetup{colorlinks=true}
\makeatletter
\def\maketitle{
\@author@finish
\title@column\titleblock@produce
\suppressfloats[t]
}
\makeatother

\makeatletter

\newcommand*{\wideboxed}[1]{\setlength{\fboxsep}{1ex}%
  \fbox{\m@th$\displaystyle#1$}}
\makeatother

\begin{document}

\title{Diagnosing 2D symmetry protected topological states via mixed state anomaly}
\author{Chao Xu}
\thanks{These authors contributed equally.}
\affiliation{Institute for Advanced Study, Tsinghua University, Beijing 100084, China}
\author{Yunlong Zang}
\thanks{These authors contributed equally.}
\affiliation{Kavli Institute for Theoretical Sciences, University of Chinese Academy of Sciences, Beijing 100190, China}
\author{Yixin Ma}
\thanks{These authors contributed equally.}
\affiliation{Kavli Institute for Theoretical Sciences, University of Chinese Academy of Sciences, Beijing 100190, China}
\author{Yingfei Gu}
\email{guyingfei@tsinghua.edu.cn}
\affiliation{Institute for Advanced Study, Tsinghua University, Beijing 100084, China}
\author{Shenghan Jiang}
\email{jiangsh@ucas.ac.cn}
\affiliation{Kavli Institute for Theoretical Sciences, University of Chinese Academy of Sciences, Beijing 100190, China}
\date{\today}

\date{\today}
\begin{abstract}
Symmetry-protected topological (SPT) phases are short-range entangled quantum states characterized by anomalous edge behavior, a manifestation of the bulk-boundary correspondence for topological phases.  
Moreover, the Li-Haldane conjecture posits that the entanglement spectrum exhibits the same anomaly as the physical edge spectrum, thereby serving as an entanglement-based fingerprint for identifying topological phases.  
In this work, we extend the entanglement-based diagnostic tools by demonstrating that the edge anomaly is manifested not only in the entanglement spectrum but also in the reduced density matrix itself, a phenomenon we refer to as the \emph{mixed state anomaly}.  
Focusing on the two-dimensional $\mathbb{Z}_2$ SPT phase, we show that this anomaly is subtly encoded in symmetry-twisted mixed states, leading to a topological contribution to the disorder parameter beyond the area law,  as well as a spontaneous-symmetry-breaking type long-range order when time reversal symmetry is present. 
\end{abstract}

\maketitle

\emph{Introduction.}  
Symmetry and topology have been proven to be fruitful guiding principles in the discovery and understanding of quantum many-body systems. One of the major achievements in this direction is the construction and classification of symmetry-protected topological (SPT) phases~\cite{spt_GuWen2009,spt_ChenGuWen2011,spt_chenSymmetryProtectedTopological2013,spt_senthil2015symmetry}. 
Nontrivial SPT states refer to short-range entangled states that cannot be transformed into trivial product states using symmetric finite-depth local unitaries.
These states exhibit featureless bulk excitations yet support anomalous edge states. 
As an example, in 2D spin systems with $\ZZ_2$ Ising symmetry, there are two classes of SPT phases: one is the conventional Ising paramagnet, and the other is the topological Ising paramagnet, which hosts symmetry-breaking or gapless edge modes~\cite{spt_chen_mpo, spt_levin2012braiding,chen-wen2012}.  

To diagnose SPT phases, one may introduce a physical boundary to investigate their anomalous edge states~\cite{sptedge_else2014classifying,sptedge_wang2015bosonic,sptedge_bultinck2018global,sptedge_kawagoe2021anomalies}, or couple the system to gauge fields for further analysis~\cite{gauging_Ye2013symmetry,gauging_Wen2014gauging,gauging_zaletel2014detecting,gauging_cheng2014topological,gauging_wang2015field}.
Both approaches are ``Hamiltonian-based'', requiring specific modifications to the physical systems. 
Alternatively, ``wavefunction-based'' methods such as strange correlators~\cite{strangecorr_you2014wave} and wavefunction surgeries~\cite{pollmann2012,mbti_shapourian2017many,mbti_Ken2017many,mbti_Ken2018many,turzillo2025detection} are also known, which make use of the entire wavefunction. 

In this work, we utilize the notion of the ``entanglement Hilbert space''~\cite{fidkowski2011topological} to detect SPT phases, based on the anomalous symmetry action on mixed states in such space. 
Instead of requiring the entire wavefunction, this framework only necessitates the reduced density matrix (RDM) of the target state on local patches of size larger than the correlation length. 
Moreover, the entanglement cuts used to define these patches do not require physical edges and are therefore independent of the details of edge Hamiltonians.
The general idea that topological properties are encoded in entanglement structures can be traced back to the concept of topological entanglement entropy~\cite{cft_kitaevPreskill, levin2006detecting}. 
Subsequently, Li and Haldane~\cite{cft_lihaldane2008} demonstrated that the spectrum of the RDM, namely the entanglement spectrum, contains more information than a single number such as the entanglement entropy.
Physically, the entanglement spectrum is arguably related to edge spectrum, a manifestation of the bulk--boundary correspondence~\cite{pollmann2010entanglement,cirac2011entanglement,qi2012general}.

Our approach further extends the bulk--boundary correspondence from the entanglement spectrum to the RDM itself, revealing that the latter contains even richer topological information~\cite{modularC,modularCfree}.
Focusing on the 2D $\ZZ_2$ SPT case, we demonstrate that the RDM, when expressed in the entanglement Hilbert space, encodes the characteristic edge anomaly.
By introducing the symmetry-twisted RDM~\cite{diso_zang2024detecting,cai2025universal}, we detect this anomaly by the protected degeneracies in the correlation spectrum.

\emph{The setup.} 
Consider a ground state wavefunction $\ket{\psi}$ of the 2D bosonic SPT protected by a global $\ZZ_2$ symmetry $G=\{1,g\}$, where $g$ is represented as $U(g)= \bigotimes_j U_j(g) $ with $U_j(g)$ acting on site $j$.
We partition the system into a disk-shaped region $A$ and its complement $\bar{A}$. 
The RDM for region $A$ is
\begin{equation}
    \adjincludegraphics[valign=c]{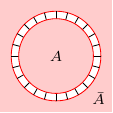}~,
    \quad
    \rho_A=\Tr_{\bar{A}}\ket{\psi}\bra{\psi}=\sum_{\alpha=1}^{\chi_A}\ee^{-\lambda_\alpha}\ket{\phi_A^\alpha}\bra{\phi_A^\alpha}
    \label{eq:rdm}
\end{equation}
where the Schmidt index $\alpha$ naturally defines a $\chi_{A}$ dimensional \emph{entanglement Hilbert space}, isomorphic to the subspace $\mathrm{span}\{\ket{\phi^\alpha_A}|_{\alpha=1,\dots,\chi_A}\}$.
For the $\ZZ_2$ SPT, we expect the entanglement Hilbert space can be approximated by a tensor-factorized Hilbert space $\HH_{\partial A}=\bigotimes_{j=1}^{L_{\partial A}} \HH_j$ localized at the 1D entanglement-cut $\partial A$ (as indicated by black bars in the above cartoon), consistent with the entanglement area-law 
\footnote{The approximation of the entanglement Hilbert space as a tensor product of local Hilbert spaces on the one-dimensional cut is a sufficient, though not necessary, condition for the area-law entanglement entropy. 
For example, topological phases with \emph{chiral edge modes} exhibit area-law entanglement entropy, yet their entanglement Hilbert spaces are generally believed \emph{not} to admit a tensor-product structure associated with a one-dimensional lattice.}.
As a concrete example, for a projected entangled pair state~(PEPS), $\HH_{j}$ corresponds to virtual leg $j$ at boundary $\partial A$~\cite{verstraete2004renormalization}. 
The dimension reduction from the bulk Hilbert space $\HH_A$ to the entanglement Hilbert space at 1D cut induces an isomorphism from the 2D RDM $\rho_A$ to a 1D mixed state $\rho_{\partial A}$ on $\HH_{\partial A}$~\cite{cirac2011entanglement,Sun2024holographic}, also see supplemental section A of the supplemental material~\cite{SM} for details.

A key feature of SPT phases within the entanglement bulk-boundary correspondence framework is that, although $\HH_{\partial A}$ factorizes, the induced symmetry action --- denoted by $W_{\partial A}(g)$ --- does \emph{not} factorize trivially into local actions on each site $\mathcal{H}_j$.  
This implies that $\rho_{\partial A}$ realizes a mixed state with an anomalous global symmetry, a phenomenon we refer to as the \emph{mixed state anomaly}\cite{HuangSunDiehl2022topological,Kawabata2024LSM,Wang2025anomaly,hsin2024anomalies,Huang2025interaction}.
We begin by demonstrating the central consequence of this anomaly, followed by concrete examples, heuristic arguments from field theory, and a more rigorous analysis using tensor network methods.

\begin{figure}[t]
    \includegraphics{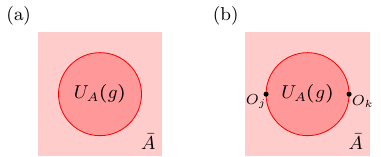}
    \caption{Diagnostics proposed for detecting 2D $\ZZ_2$ SPT: (a) disorder parameter $\lrangle{U_A(g)}$; (b) The $\ZZ_2$ charge correlator at the boundary of the disorder operator, denoted as $\lrangle{O_jO_k}_A^g$.}
    \label{fig:dis_IMO}
\end{figure}

\emph{Main results.} 
We claim that the $\ZZ_2$ disorder operator $U_A(g)=\otimes_{j\in A}U_j(g)$ on a large patch $A$ (depicted in Fig.~(\ref{fig:dis_IMO}a)) has the following expectation value (also known as disorder parameter) on the ground state $|\psi\rangle$ of a $\ZZ_2$ SPT
\begin{equation}
\wideboxed{
    -\ln\lrangle{U_A(g)}_{\psi}
    =\alpha L_{\partial A}- \underbrace{\ln  D   }_{\rm top.}- \ln \cos(\theta L_{\partial A}) +\dots,
    }
    \label{eq:disorder_para}
\end{equation}
where $L_{\partial A}$ is the boundary length of region $A$, $\alpha$ is a non-universal coefficient, $D\geq 2$ is an integer that implies the non-triviality of the $\ZZ_2$ SPT.
We note that the above form for disorder parameter first appeared in Ref.~\cite{cai2025universal}, in the context of the thermal states of SPT edges. 

A few comments are in order 
(1) The leading area-law term reflects the symmetry-preserving nature of the SPT phase, in contrast to the symmetry-breaking states where a volume law is expected-- this term was the original motivation for the disorder operator~\cite{diso_kadanoff1971determination}. 
(2) Disorder parameters have emerged as powerful tools for probing non-trivial phases and phase transitions across diverse systems, including crystalline SPT phases~\cite{pollmann2012,mbti_Ken2017many,Barkeshli2023}, symmetry-enriched topological orders~\cite{diso_chen2022topological}, superfluids~\cite{diso_wang2024distinguish}, gapless systems~\cite{Wu2021universal, chengmeng_MC2021, chengmeng_MC2021b, chengmeng_MC2022, chengmeng_MC2023, cai2024disorderoperators2dfermi,huang_xu_2025}, and certain mixed state phases with anomalous features~\cite{cai2025universal,diso_zang2024detecting}.
(3) An oscillatory sub-leading term may also arise, with $\theta$ being non-universal phase \cite{cai2025universal}. 
(4) The disorder parameter may be interpreted as the trace of a \emph{symmetry twisted density matrix} $\rho_A^g:=\rho_A U_A(g)$, where $\Tr(\rho_A^g) =\lrangle{U_A(g)}$. 
Through the entanglement bulk-boundary correspondence mentioned in the setup section, $\rho_A^g$ reduces to its 1D counterpart $\rho_{\partial A}^g := \rho_{\partial A} W_{\partial A}(g)$.
Then the integer $D$ in Eq.~\eqref{eq:disorder_para} corresponds precisely to the degeneracy of the dominant eigenvalues in the transfer matrix spectrum for $\rho_{\partial A}^g$ - a manifestation of the anomalous symmetry action on the mixed state. 
(5) When the system has additional symmetries, the degeneracy $D$ may exceed $2$.
A notable case occurs when time-reversal symmetry is present, as in the Levin-Gu wavefunction that will be discussed momentarily, we find $D=4$. 
Remarkably, in such case, the dominant eigenvectors of the transfer matrix for $\rho_{\partial A}^g$ carry distinct $\ZZ_2$ quantum numbers, which can be probed through long-range correlation of $\ZZ_2$ charged operators at the boundary of the disorder operator~(depicted in Fig.~(\ref{fig:dis_IMO}b)):
\begin{equation}
\wideboxed{
    \langle O_j O_k \rangle_{A}^g  := \frac{\lrangle{U_A(g)O_j O_k}_\psi}{\lrangle{U_A(g)}_\psi} 
    =\frac{\Tr(\rho_{\partial A}^g O_j O_k)}{\Tr(\rho_{\partial A}^g)}
    \sim  O(1)}
     \label{eq:imo_correlator}
\end{equation}
where $O_{j/k}$ are $\ZZ_2$ charged operators that are localized near $\partial A$ but far away from each other.

\emph{Example (Levin-Gu state).} 
Now, we demonstrate explicitly Eq.~\eqref{eq:disorder_para} and Eq.~\eqref{eq:imo_correlator} on the Levin-Gu (LG) state  $\ket{\Psi^{\rm{LG}}} $, a fixed point wavefunction of the 2D $\ZZ_2$ SPT phase \cite{spt_levin2012braiding}. 
It is defined on a triangular lattice of Ising spins as a signed superposition of all spin configurations:
\begin{equation}
    \ket{\Psi^{\rm{LG}}} = \frac{1}{2^{N/2}}\sum_{\text{all spin configs.}}(-1)^{N_{dw}(c)}\ket{c} .
\end{equation} 
where $N_{dw}(c)$ counts the number of domain walls in spin configuration $c$. 
The LG state can be generated by a finite depth unitary circuit, denoted as $U^{\rm LG}$, acting on a direct product state $\ket{\Omega^+}$: 
\begin{equation}
    \ket{\Psi^{\rm{LG}}} = \underbrace{ \prod_{f\in\rm faces}  {\rm CC} Z_{f} 
 \prod_{e\in\rm edges}  {\rm C}Z_{e}
 \prod_{v\in\rm vertices}  Z_{v}}_{=:U^{\rm LG}} 
 \underbrace{\ket{+}^{\otimes N}}_{=: \ket{\Omega^{+}}}
\end{equation} 
Here, we have assigned a Pauli-$Z$ to each vertex, control-$Z$ to each edge and control-control-$Z$ to each face. 
The global $\ZZ_2$ symmetry in question is generated by $U^X: = \prod_{v\in\rm vertices} X_v$. 
Then the expectation value of the symmetry restricted on $A$, i.e. $U_A^X=\prod_{v\in A} X_v$ is given as follows
\begin{equation}
     \langle U^X_A \rangle =
     \bra{\Psi^{\rm{LG}}} U^X_A \ket{ \Psi^{\rm{LG}}}
     = \bra{ \Omega^{+} } (U^{\rm LG})^\dagger U^X_A U^{\rm LG} \ket{\Omega^{+} } . 
     \label{eq:lg_disorder}
\end{equation}
Since $U^X_A\ket{\Omega^{+}} = \ket{\Omega^{+}}$, the disorder parameter then equals the commutator between $U_A^X$ and $U^{\rm LG}$, whose non-trivial part is expected to be localized near the boundary. 

Indeed, explicit calculation (presented in supplemental section B) shows that the $U_A^X$ for the lower half plane
\begin{align}
    U_A^X=\adjincludegraphics[valign=c]{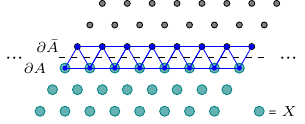}
\end{align}
has the following commutator with $U^{\rm LG}$:
\begin{equation}
    (U_A^X )^\dagger (U^{\rm LG})^\dagger U_A^X U^{\rm LG}= \prod_{e} {\rm C}Z_e \prod_{v} Z_v
\end{equation}
where the edges $e$ and vertices $v$ are taken near the boundary, denoted as the blue truss marked in the above lattice.
Therefore, Eq.~\eqref{eq:lg_disorder} reduces to the expectation value of a 1D operator. 
Explicit calculation yields 
\begin{equation}
    -\ln \lrangle{U_A^X} = \ln 2\cdot L_{\partial A} - \ln (4\cos (\pi L_{\partial A}/3))\,,
\end{equation}
with $L_{\partial A}$ the length of the boundary. 
Compared with Eq.~\eqref{eq:disorder_para}, we have $\theta=\pi/3$ and $D=4$.
Note that $D>2$ is due to the presence of additional time-reversal symmetry.

As commented previously, time-reversal symmetry induces an extra observable on the twisted density matrix
\begin{align}
    \lrangle{Z_j Z_{k}}_{A}^g = \frac{1}{3} + \frac{2}{3}\cos \frac{2\pi (j-k) }{3} +\frac{2}{3}\tan \frac{\pi L_{\partial A}}{3}\sin \frac{2\pi (j-k)}{3}\,,
\end{align}
which matches the form in Eq.~\eqref{eq:imo_correlator}. See supplemental section B and C for details.

\emph{Heuristic CFT argument.} 
For short-range correlated SPT states, a heuristic but nontrivial assumption is that its RDM  can be approximated by a high temperature conformal field theory (CFT), $\rho_A \propto \ee^{-\beta H_{\rm CFT}}$~\cite{cft_kitaevPreskill,cft_lihaldane2008}.
More specifically, for the $\ZZ_2$ SPT in question, the representative entanglement Hamiltonian $H_{\rm CFT}$ has been identified as the free boson CFT \cite{spt_levin2012braiding,cft_chen2012_chiral,cft_Azses2023}:
\begin{equation}
    H_{\rm CFT}=\frac{1}{8\pi}\int_0^L {\rm d}x\big(R^2(\partial_x \phi)^2 + \tilde{R}^2(\partial_x \theta)^2\big).
     \label{eq:free_boson}
\end{equation}

Generalizing Cardy's argument~\cite{cft_cardy1986}, under modular transformation, the high temperature (i.e. $\beta \sim \Delta^{-1} \ll L$, with $L=L_{\partial A}$) partition function with a symmetry twist is transformed to a low temperature one with a flux line inserted: 
\begin{equation*}
\adjincludegraphics[valign=c]{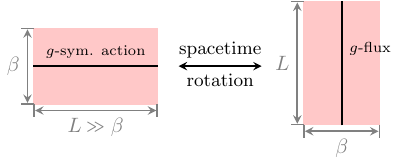}
\end{equation*}
On the right hand side,  the $\ZZ_2$-flux effectively generates an anti-periodic boundary condition for the Hamiltonian, denoted as $H_{\rm CFT}(g)$.
For free boson theory in Eq.~\eqref{eq:free_boson}, $H_{\rm CFT}(g)$ has 4-fold ground state degeneracy \cite{cft_affleck1992,cft_clay2020}. So the disorder parameter yields
\begin{equation}
    \lrangle{U_A(g)}\sim 4\ee^{-L {E}_0(g)} +\cdots,
\end{equation}
which matches Eq.~\eqref{eq:disorder_para}.
$D=4$ here is consistent with the fact that Eq.~\eqref{eq:free_boson} is time-reversal symmetric.

We remark here that lattice models are generally more complicated than their continuum field theory counterparts -- particularly due to the absence of explicit space-time symmetry on the lattice.
For instance, as demonstrated in supplemental section D, UV-complete lattice realizations of the $\ZZ_2$ SPT edge can produce ground state degeneracies that differ from the field-theoretic predictions, thereby invalidates the heuristic argument.
Given these subtleties, we now turn to a more rigorous treatment based on tensor network formalism that is native on lattice.

\emph{RDM as a 1D anomalous mixed state.}
As motivated in the setup, the RDM $\rho_A$ of a $\ZZ_2$ SPT state $\ket{\psi}$ is isomorphic to a 1D mixed state $\rho_{\partial A}$ on the entanglement Hilbert space $\HH_{\partial A}$. 
The global symmetry $U(g)$ of the pure state $\ket{\psi}$ induces a weak symmetry $U_A(g)$ on the RDM $\rho_A$. 
As detailed in supplemental section A, through the entanglement bulk-boundary correspondence, this translates to a weak symmetry $W_{\partial A}(g)$ on $\rho_{\partial A}$:
\begin{align}
    \rho_{\partial A}=W_{\partial A}(g)\cdot\rho_{\partial A}\cdot W^\dg_{\partial A}(g).
    \label{eq:rho_bdry_weak_sym}
\end{align}
Crucially, for the $\ZZ_2$ SPT phase, $W_{\partial A}(g)$ does not factorize into a product of on-site $\ZZ_2$ symmetry action, sharing the same anomaly for a physical edge \cite{Williamson2016MPO,spt_jiang2016}.
To properly represent this anomaly, two approaches are available:
\begin{enumerate}
    \item Non-onsite realization: for instance, express $W_{\partial A}$ as a matrix product $\ZZ_2$ operation~\cite{spt_chen_mpo,sptedge_else2014classifying,Williamson2016MPO};
    \item Extend the $\ZZ_2$ symmetry group: enlarge the entanglement Hilbert space with an extended onsite symmetry group~(e.g. $\ZZ_4$), and then recover the anomalous $\ZZ_2$ action through appropriate projection~\cite{spt_jiang2016,wang2018symmetric}.\label{item:group_ext}
\end{enumerate}
In this work, we adopt the second approach, as it facilitates a clear derivation of our central result in Eq.~\eqref{eq:disorder_para} and Eq.~\eqref{eq:imo_correlator}.

To see how the approach \eqref{item:group_ext} works explicitly, let us start with a minimal example that realizes the $\ZZ_2\rightarrow \ZZ_4$ extension:  
let each site at the entanglement cut contains two qubits, namely $\HH_j=(\mathbb{C}^2)_l\otimes(\mathbb{C}^2)_r$, and implement a $\ZZ_4$ onsite symmetry as $W_{\partial A}=\prod_{j\in\partial A} W_j(g)$, with $W_j(g)=X_{j}^lX_{j}^r\exp ({\ii\frac{\pi}{4}(1-Z_{j}^lZ_{j}^r)})$.
Now, we define the effective (or restricted) entanglement Hilbert space by the projector $P=\prod_j\frac{1}{2}(1+Z_{j}^rZ_{j+1}^l)$.
Hence, in the projected space, $W(g)$ becomes $\tilde{W}(g)=\big( \prod_j \tilde{X}_{j} \big) \cdot \exp\big( \ii\frac{\pi}{4}\sum_j(1-\tilde{Z}_{j}\tilde{Z}_{j+1}) \big)$, with identification $\tilde{X}_{j}:= X_{j}^r X_{j+1}^l$, and $\tilde{Z}_{j}:= Z_{j}^r=Z_{j+1}^l$ serving as the Pauli operators on the effective qubits.
This procedure reproduces the anomalous $\ZZ_2$ symmetry for the edge theory of the Levin-Gu state~\cite{spt_chen_mpo,spt_levin2012braiding}.
Physically, the projector $P$ defines a low energy space where the so-called 1D $\ZZ_4$ intrinsic gapless SPT is realized~\cite{gspt_thorngren2021intrinsically, gspt_wen2023bulk,gspt_li2024decorated}.

In more general settings, the local Hilbert space $\HH_j$ could be distinct from the minimal example, yet the operator algebra is preserved.
More explicitly, the Pauli operators $Z^{l/r}$ are generalized to two commuting operators $\lambda^{l/r}$ satisfying $(\lambda^{l/r} )^2=1$.
The effective entanglement space is defined through projector $P:=\prod_j\frac{1}{2}(1+\lambda_{j}^r\lambda_{j+1}^l)$, which naturally gives the following constraints on $\rho_{\partial A}$:
\begin{equation}
    \rho_{\partial A}
    =\lambda_{j}^r\lambda_{j+1}^l\cdot \rho_{\partial A}
    =\rho_{\partial A}\cdot \lambda_{j}^r\lambda_{j+1}^l
    \label{eq:rho_bdry_plq_igg}
\end{equation}
Note that the product of these constraints -- namely $D_{\partial A}(g):=\prod_{j\in\partial A} \lambda_j^r \lambda_{j+1}^l$ -- may be interpreted as a strong $\ZZ_2$ symmetry, while $W_{\partial A}(g)$ generates an onsite weak $\ZZ_4$ symmetry on $\rho_{\partial A}$. 
These operators satisfy the following algebraic relations~\cite{spt_jiang2016}:
\begin{equation}
    \begin{aligned}
     &[W_j(g)]^2=D_j\equiv\lambda_j^l\cdot \lambda_j^r~,\\
    &W_j(g)\cdot\lambda_j^{l/r}=-\lambda_j^{l/r}\cdot W_j(g)
    \end{aligned}
    \label{eq:z2_spt_tensor_eq}
\end{equation}

The above framework also provides generic PEPS construction for the $\ZZ_2$ SPT phase, offering a bulk interpretation of anyon condensation~\cite{spt_jiang2016}.
The strong $\ZZ_2$ symmetry ($D_{\partial A}$) originates from a $\ZZ_2$ virtual-leg symmetry in PEPS tensors, which corresponds to an $m$-anyon loop of the Toric Code topological order~\cite{schuch2010peps}~(see supplemental section A). 
For an open string $M \in \partial A$, the operator $W_M(g):=\prod_{j\in M}W_j(g)$ creates $g$-flux line terminating at $g$-defects.
Eq.~\eqref{eq:z2_spt_tensor_eq} implies the fusion rule $[W_M(g)]^2=D_M$, indicating that two $g$-defects fuse into an $m$-anyon. 
When the decomposisition $D_j=\lambda_j^l\lambda_j^r$ is imposed with $\lambda^{l/r}$ carrying $g$-charge (from the second line of Eq.~\eqref{eq:z2_spt_tensor_eq}), Eq.~\eqref{eq:rho_bdry_plq_igg} reveals that the tensor wavefunction is invariant under $\lambda_1^l\cdot D_M\cdot \lambda_n^r$ on virtual legs.
This signals a condensation of bound states of $m$-anyon and $g$-charge, 
resulting a
$\ZZ_2$ SPT phase~\cite{Wang2013weak,ellison2022pauli}.

Having established the anomalous symmetry on the entanglement Hilbert space $\HH_{\partial A}$, we are ready to study the mixed state $\rho_{\partial A}$ wherein. 
$\rho_{\partial A}$ hosts the following properties: 
(1) As $\rho_{\partial A}$ is the RDM of an SPT state, all correlators in $\rho_{\partial A}$ decay exponentially;
(2) $\rho_{\partial A}$ possesses a weak $\ZZ_4$ symmetry generated by $W_{\partial A}(g)$, and a strong $\ZZ_2$ subgroup symmetry generated by $[W_{\partial A}(g)]^2=D_{\partial A}$;
(3) The projection constraint~\eqref{eq:rho_bdry_plq_igg}  implies a string order parameter, 
\begin{equation}
    \lim_{\abs{l}\to\infty}\Tr\big[ \rho_{\partial A}\cdot \big(\lambda_1^r\cdot \big( \prod_{j=2}^{l}D_{j} \big)\cdot \lambda_{l+1}^l\big) \big]=1
\end{equation}    
which is identified as a $\ZZ_2$ strong symmetry string decorated with double-$\ZZ_4$-charged endpoints. 
We remark that these properties identify $\rho_{\partial A}$ belongs to the recent proposed 1D intrinsic average $\ZZ_4$ SPT phase~\cite{mix_ma2023intrinsic}, but without disorder or decoherence.

In supplemental section E, we generalize Eq.~\eqref{eq:z2_spt_tensor_eq}, Eq.~\eqref{eq:rho_bdry_weak_sym}, and Eq.~\eqref{eq:rho_bdry_plq_igg} to 2D SPT phases classified by group cohomology $H^3(G,U(1))$ for arbitrary finite group $G$.
Furthermore, using the mathematical framework of crossed module, we establish general correspondence between 2D SPT phases and 1D intrinsic average SPT phases
\footnote{We note that Ref.~\cite{Sun2024holographic} proposes a similar mapping between $(d+1)$-dimensional subsystem SPT states and $d$-dimensional intrinsic average SPT phases.
    However, the origin of the strong symmetry differs fundamentally:
    In their framework, it arises directly from the subsystem symmetry of the physical Hamiltonian, while in our work, it emerges from the edge anomaly rather than any physical symmetry.}.

\emph{Observables from anomalous RDM.}
We now proceed to sketch (full details are shown in supplemental section~F \cite{SM}) our derivation of the main results Eq.~\eqref{eq:disorder_para} and Eq.~\eqref{eq:imo_correlator} based on properties of $\rho_{\partial A}$. 
We assume $\rho_{\partial A}$ to be translation-invariant and representable as a Matrix Product Density Operator~(MPDO):
\begin{equation}
    \rho_{\partial A} = \Tr \left[M^{L_{\partial A}} \right]
    ~=~\adjincludegraphics[valign=c]{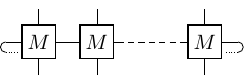}\,,
\end{equation}
where $M$ denotes the local tensor of the MPDO.
Hence, the disorder parameter reads
\begin{equation}
    \begin{aligned}
        \lrangle{U_A(g)}=\Tr(\underbrace{\rho_{\partial A}W_{\partial A}(g)}_{=:\rho_{\partial A}^g})
        = \adjincludegraphics[valign=c]{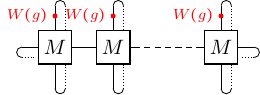}
    \end{aligned}
    \label{eq:trans_matrix_def}
\end{equation}
namely, the trace of the symmetry twisted density matrix $\rho^g_{\partial A}$ can be expressed by the symmetry-twisted transfer matrix
$ T(g) := \adjincludegraphics[valign=c]{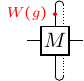}\,. $
Consequently, $\lrangle{U_A(g)}$ is determined by the dominant eigenvalues of $T(g)$ when $L_{\partial A}\to\infty$.

Next, we show that the spectrum of $T(g)$ exhibits degeneracy due to the anomalous symmetry action, leading to the topological term in the disorder parameter.
Symmetry constraints in Eq.~\eqref{eq:rho_bdry_weak_sym} and Eq.~\eqref{eq:rho_bdry_plq_igg} induce gauge transformations on the internal lines of $M$:
\begin{align}
    \adjincludegraphics[valign=c]{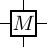} = {}& \adjincludegraphics[valign=c]{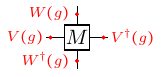} = \adjincludegraphics[valign=c]{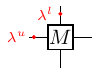} = \adjincludegraphics[valign=c]{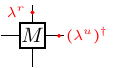}\nonumber \\
    ={}&\adjincludegraphics[valign=c]{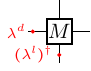} = \adjincludegraphics[valign=c]{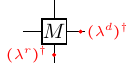} 
\end{align}
where algebraic relations between $V(g)$ and $\lambda^{u/d}$ follows Eq.~\eqref{eq:z2_spt_tensor_eq}:
\begin{equation}
    \begin{aligned}
        &[V(g)]^2=\lambda^u\lambda^d~,\quad
        [\lambda^{u/d}]^2=1\\
        &V(g)\cdot\lambda^{u/d}=-\lambda^{u/d}\cdot V(g)\\
    \end{aligned}
    \label{eq:rho_vleg_tensor_eq}
\end{equation}

We then impose Hermiticity condition $\rho_{\partial A}=\rho_{\partial A}^\dg$, which can be viewed as an anti-unitary $\ZZ_2$ symmetry that interchanges the ket and bra space. 
Thus, it imposes an extra constraint on $M$:
\begin{align}
    \adjincludegraphics[valign=c]{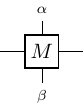} = \adjincludegraphics[valign=c]{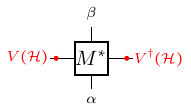}
    \label{eq:rho_local_tensor_hermitian}
\end{align}
where $V(\HH)\KK$ commutes with $V(g)$, and $V(\HH)\KK\cdot \lambda^u\cdot (V(\HH)\KK)^\dg=\lambda^d$.
It is then straightforward to derive that $T(g)$ is invariant under anti-unitary operation $V(\mathcal{H})\mathcal{K}\lambda^u$. 
As $(V(\HH)\KK\lambda^u)^2\cdot T(g) =\lambda^u\lambda^r\cdot T(g)= -T(g)$, spectrum of $T(g)$ are pairs of Kramers doublets, whose eigenvalues are conjugate to each other.

Denoting the dominant eigenvalues of $T(g)$ as $\exp(-\alpha\pm\ii\theta)$, for large $L_{\partial A}$, the disorder parameter is then given by
\begin{equation}
    -\ln\lrangle{U_A(g)}=\alpha\cdot L_{\partial A}-\ln2-\ln\cos(\theta L_{\partial A})+\dots
    \label{}
\end{equation}
which contains the promised topological term beyond the conventional area law. 

Next, we show that additional time-reversal symmetry $\TT$ could further shift the topological term, as well as give rise to a spontaneous-symmetry-breaking–type long-range order.
Time reversal symmetry imposes the following local tensor constraint:
\begin{equation}
    \adjincludegraphics[valign=c]{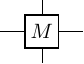} = \adjincludegraphics[valign=c]{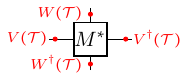}
\end{equation}
Here, we consider the case where $W(\TT)\KK$ commutes with $W(g)$, which also implies $V(\TT)\KK$ and $V(g)$ commute.

From Eq.~\eqref{eq:rho_vleg_tensor_eq}, we have $[V(g)]^2\cdot T(g)=-T(g)$. 
Therefore, for an arbitrary eigenstate of $T(g)$, denoted as $\vket{\psi}$, we have $V(g)\vket{\psi} = \pm \ii\vket{\psi}$.
As the anti-unitary symmetry $V(\TT)\KK$ commutes with $V(g)$, $V(\TT)\KK\vket{\psi}$ carries opposite $V(g)$ quantum number from $\vket{\psi}$\footnote{In contrast, the anti-unitary symmetry $V(\HH)\KK\lambda^u$ anti-commutes with $V(g)$, so $V(\HH)\KK\lambda^u\vket{\psi}$ and $\vket{\psi}$ share the same $V(g)$ quantum number.}.
Therefore, with additional $\TT$ symmetry, the spectrum of $T(g)$ is at least four fold degeneracy in terms of modulus, and can be partitioned into two pairs carrying opposite $V(g)$ quantum number $\pm\ii$. 
Such degeneracy gives $D=4$ in Eq.~\eqref{eq:disorder_para}.
These degenerate eigenstates with opposite $V(g)$ quantum number lead to further predictions as Eq.~\eqref{eq:imo_correlator}: for local operator $O$ carrying $g$-charge, we have~\cite{SM}
\begin{equation}
    \frac{\lrangle{U_A(g)O_j O_k}}{\lrangle{U_A(g)}}\xrightarrow{\abs{j-k}\to\infty} O(1)
     \label{}
\end{equation}
where $j,k$ are sites near $\partial A$
\footnote{It should be noted that our prediction works for $O$ acting on the entanglement space.
Nonetheless, it is reasonable to argue that the present result remains valid for $\ZZ_2$ charged operators acting on the physical legs in the vicinity of the entanglement cut.}.

\emph{Summary and discussion.}
This work reveals a deep connection between topological phases of matter and mixed-state quantum phases.
By extending the bulk-boundary correspondence, we demonstrate that the RDM of a 2D SPT phase, when expressed in the entanglement space, becomes a 1D mixed state encoding the characteristic quantum anomaly.
Focusing on the $\ZZ_2$ SPT phase, we show that quantum anomaly of the 1D RDM manifests as an intrinsic average $\ZZ_4$ SPT phase, where the $\ZZ_4$ weak symmetry is extended from the global $\ZZ_2$ symmetry.
These anomalies produce measurable signatures including a robust topological term in the disorder parameter \eqref{eq:disorder_para} and long-range order in twisted mixed state \eqref{eq:imo_correlator} when time-reversal symmetry is present.
Our framework generalizes beyond the previous Lieb-Schultz-Mattis-type anomalies in mixed states~\cite{Zhou_2024,diso_zang2024detecting} to encompass broader 't Hooft anomalies.

The RDM perspective further enables new approaches for detecting topological phases with anyonic excitations \cite{Zang}, while insights from topological phases conversely inform the study of nontrivial mixed-state phases.
This two-way connection between topological phases and mixed-state phases could benefit other problems, such as classification and detection of mixed-state phases \cite{BudichDiehl2015topology,Bardyn2018probing,Altland2021symmetry,Sieberer2025universality,HuangDiehl2024mixed,lu2024bilayer,Ellison2025toward,mao2024probingtopologygaussianmixed}, as well as phase transitions between mixed states~\cite{Lee2023quantumcriticality,Guo2025SWSSB,ando2024gaugetheorymixedstate}.

\emph{Acknowledgment.}
We thank Ruochen Ma, Meng Cheng, Yunqin Zheng, and Xie Chen for helpful discussions.
We are especially grateful to Meng Cheng who read a preliminary version of the manuscript and made some valuable comments.
This work is supported by MOST NO. 2022YFA1403902, NSFC Grant No.12342501, and the National Key R\&D Program of China 2023YFA1406702. Y.G. is also supported by the Tsinghua University Dushi program and the DAMO Academy Young Fellow program.

\emph{Note added.}
Upon completing this work, we became aware of a related paper by Sala, Pollmann, Oshikawa, and You~\cite{sala2025entanglementholographyquantumphases} that also examines the anomalous symmetries on RDM of SPT phases. 

\bibliography{lg.bib}
\end{document}


\title{Supplemental Materials: Diagnosing 2D symmetry protected topological states via mixed state anomaly}
\author{Chao Xu}
\thanks{These authors contributed equally.}
\affiliation{Institute for Advanced Study, Tsinghua University, Beijing 100084, China}
\author{Yunlong Zang}
\thanks{These authors contributed equally.}
\affiliation{Kavli Institute for Theoretical Sciences, University of Chinese Academy of Sciences, Beijing 100190, China}
\author{Yixin Ma}
\thanks{These authors contributed equally.}
\affiliation{Kavli Institute for Theoretical Sciences, University of Chinese Academy of Sciences, Beijing 100190, China}
\author{Yingfei Gu}
\email{guyingfei@tsinghua.edu.cn}
\affiliation{Institute for Advanced Study, Tsinghua University, Beijing 100084, China}
\author{Shenghan Jiang}
\email{jiangsh@ucas.ac.cn}
\affiliation{Kavli Institute for Theoretical Sciences, University of Chinese Academy of Sciences, Beijing 100190, China}
\date{\today}

\maketitle

\tableofcontents

\hfill

In this supplemental material, we present a brief explanation of symmetries of reduced density matrix~(RDM) on the entanglement space (Sec.~\ref{sec:rdm_sym}), a detailed calculation of disorder operator on Levin-Gu state based on quantum circuits (Sec.~\ref{app:LG_fp}) or the entanglement space RDM (Sec.~\ref{sec:mpdo_lg}), the numerical calculation of GSD of a 1D lattice realization of the edge theory of the $\ZZ_2$ SPT phase (Sec.~\ref{sec:edge_lg_z4}), the RDM for generic 2D SPT states (Sec.~\ref{sec:rdm_spt}), and the details of MPDO-based proof (Sec.~\ref{sec:mpdo_proof}).

\section{RDM on the entanglement space and its symmetries}
\label{sec:rdm_sym}
Consider a state $\ket{\psi}$ with a spatial bipartition into $A$ and its complement $\bar{A}$.
The Schmidt decomposition yields $\ket{\psi}= \sum^{\chi_A}_{\alpha=1} \ee^{-\lambda_{\alpha}/2} \ket{\phi_{A}^{\alpha}} \ket{\phi_{\bar{A}}^{\alpha}}$, where $\{\lambda_\alpha\}$ are entanglement spectrum, and $\chi_A$ is the Schmidt rank.
We now introduce the entanglement space $\HH_{\partial A}$ of dimension $\chi_A$, enabling a tensor representation of the Schmidt decomposition:
\begin{equation}
    \ket{\psi} \equiv \sum_{\alpha} \left( \ket{\phi_A^\alpha}\vbra{\phi_{\partial A}^\alpha} \right) \cdot \left( \ee^{-\lambda_\alpha/2} \vket{\phi_{\partial A}^\alpha}\vket{\phi_{\partial A}^\alpha} \right) \cdot \left( \vbra{\phi_{\partial A}^\alpha}\ket{\phi_{\bar{A}}^\alpha} \right)
    \equiv \Tr_{\partial A} [T_{A}\otimes\Lambda_{\partial A}\otimes T_{\bar{A}}]
    \label{eq:wf_schmit_decomp_sm}
\end{equation}
with a diagrammatic representation:
\begin{equation}
    |\psi\rangle=\begin{tikzpicture}
    [scale=0.8,baseline=-4pt]
        \draw[thick] (-0.1,0.18)-- (-0.1,0.6);
        \node at (0,-0.15) {$T_{A}$};
        \draw[thick,cyan] (0.3,-0.15)--(0.8,-0.15);
        \node at (1.3,-0.15) {$\Lambda_{\partial A}$};
        \draw[thick,cyan] (1.8,-0.15) -- (2.4,-0.15);
        \node at (2.8,-0.15) {$T_{\bar{A}}$};
        \draw[thick] (2.7,0.18)-- (2.7,0.6);
    \end{tikzpicture}.\nonumber
\end{equation}
Here, $\{\vket{\phi_{\partial A}^\alpha}\}$ is an orthonormal basis for $\HH_{\partial A}$, $T_A$ and $T_{\bar{A}}$ are isometric tensors, and $\Tr_{\partial A}$ denotes contracting states in $\HH_{\partial A}$.
We point out that, if $\ket{\psi}$ is a PEPS, $\HH_{\partial A}$ can be identified as virtual legs at $\partial A$: $\HH_{\partial A}\equiv\bigotimes_{j\in\partial A}\HH_j$.

The RDM then also admits a tensor representation: $\rho_{A}=\Tr[T_A\otimes \Lambda_{\partial_A}\otimes \Lambda_{\partial_A}^\dg\otimes T_A^\dg]$, leading to an isometry map to a mixed state in $\HH_{\partial A}$ as $\rho_{\partial A}\equiv\Lambda_{\partial A}\cdot\Lambda_{\partial A}^\dg= \sum_{\alpha}\ee^{-\lambda_\alpha}\vket{\phi_{\partial A}^\alpha}\vbra{\phi_{\partial A}^\alpha}$.

We now discuss symmetries on $\rho_{\partial A}$.
Let $\ket{\psi}$ be invariant under an onsite symmetry $g$, i.e. $(U_A(g)\otimes U_{\bar{A}}(g)) K^{s(g)}\ket{\psi}=\ket{\psi}$, where $s(g)=0/1$ for unitary/anti-unitary operation, and $U_{A}(g)$ and $U_{\bar{A}}(g)$ act on region $A$ and $\bar{A}$, respectively.
$g$ symmetry induces gauge transformation $W_{\partial A}(g)$ action on $\HH_{\partial A}$, and therefore for tensors defined in Eq.~\eqref{eq:wf_schmit_decomp_sm}, we require
\begin{equation}
    U_{s}(g)\KK^{s(g)}\cdot T_s=T_s\cdot W^\dg_{\partial A}(g)~,\quad
    \Lambda_{\partial A}=(W_{\partial A}(g)\otimes W_{\partial A}(g))K^{s(g)}\cdot \Lambda_{\partial A}
    \label{eq:schmidt_tensor_sym_sm}
\end{equation} 
where $s=A/\bar{A}$.
Therefore, $\rho_{\partial A}$ hosts \emph{a weak symmetry} under $W_{\partial A}(g)\KK^{s(g)}$:
\begin{equation}
    W_{\partial A}(g)\mathcal{K}^{s(g)}\cdot \rho_{\partial A}\cdot \mathcal{K}^{s(g)}W^{\dagger}_{\partial A}(g)=\rho_{\partial A}
    \label{eq:rho_weak_sym_sm}
\end{equation}

For topological phases, the complex entanglement patterns typically require sophisticated description of the entanglement space.
A standard approach involves first enlarging the entangelement Hilbert space, and then projecting onto a subspace that encodes the effective Schmidt states. 
This subspace is then defined by invariance under certain operation $D_{\partial A}$ on entanglement legs, yielding the following tensor constraints:
\begin{equation}
    T_s=T_s\cdot D_{\partial A}~,\quad
    \Lambda_{\partial A}=(D_{\partial A}\otimes \hat{1})\cdot \Lambda_{\partial A}
    =(1\otimes D_{\partial A})\cdot \Lambda_{\partial A}
    \label{eq:schmidt_tensor_igg_sm}
\end{equation}
Remarkably, these constraints directly imply \emph{the strong symmetry condition} for $\rho_{\partial A}$:
\begin{equation}
    \rho_{\partial A}=D_{\partial A}\cdot\rho_{\partial A}
    =\rho_{\partial A}\cdot D_{\partial A}^\dg
    \label{eq:rho_strong_sym_sm}
\end{equation}
A paradigmatic example is the PEPS representation toric code wavefunction.
Here, tensors hosts $\ZZ_2$ virtual leg symmetry: $D_{\partial A}=\prod_{j\in \partial A} D_j$ with $D_j^2=1$.
$D_{\partial A}$ is interpreted as $\ZZ_2$ flux loop~(m-anyon loop), and an open $D$ string creates $m$-anyons at its endpoints.

\section{Quantum circuits for Levin-Gu state and the disorder operator}
\label{app:LG_fp}
The Levin-Gu (LG) state\cite{spt_levin2012braiding} is the fixed point wavefunction of the 2D $\ZZ_2 = \{1,~g\}$ symmetric SPT state on a triangular lattice Ising spins.
This LG state equals a signed superposition of all Ising spin configurations
\begin{align}
    \ket{\Psi^{\rm{LG}}} = \frac{1}{2^{N/2}}\sum_{c}(-1)^{N_{dw}(c)}\ket{c}
    \label{}
\end{align}
with $c$ being Ising spin configurations, and $N_{dw}(c)$ the number of domain wall loops in the configuration $c$.
Such state can be constructed through a quantum circuit that involves non-Clifford ${\rm{CC}}Z$ gates on an initial product state $\ket{+}^{\otimes N}=\ket{\Omega^+}$:
\begin{align}
    \ket{\Psi^{\rm{LG}}}=U^{\rm{LG}}\ket{\Omega}=U^{{\rm{CC}}Z}U^{{\rm{C}}Z}U^{Z} \ket{\Omega^+} 
    \label{}
\end{align}
where
\begin{align}
    U^{{\rm{CC}}Z} = \prod_{[j,k,l]=\bigtriangleup\,\in\,\rm{faces}} U^{{\rm{CC}}Z}_{jkl};~~U^{{\rm{C}}Z} = \prod_{[j,k]\,\in\,\rm{edges}} U^{{\rm{C}}Z}_{jk};~~U^Z = \prod_{j\,\in\,\rm{vertices}} Z_j\,,
\end{align}
are the tensor product of control-control-$Z$, control-$Z$, and local Pauli-$Z$ gates on all faces, edges and sites of the lattice respectively.
The global $\ZZ_2$ symmetry is represented as
\begin{align}
    U(g) = U^X = \prod_{j}X_j\,.
\end{align}

\subsection{Disorder operator on LG state}
We now evaluate $U_A^X = \prod_{j\in A}X_j$, in which $A$ is the lower half plane of the system:
\begin{align}
    U_A^X=\begin{tikzpicture}[scale=0.9,baseline={([yshift=6pt]current bounding box.center)}]
        \usetikzlibrary{shapes.geometric}
        \foreach \x in {0,1,...,9}{
        \foreach \y in {2,...,7}{
           \filldraw[fill=black!50] (\x*10pt+\y*5pt,\y*8.66pt) circle (1pt);
        }
        }
         \foreach \x in {0,1,...,9}{
        \foreach \y in {2,...,4}{
           \filldraw[teal,fill=teal!60] (\x*10pt+\y*5pt,\y*8.66pt) circle (2pt);
        }
        }
        \draw[dashed] (4.5*5pt-5pt,4.5*8.66pt) -- (10*10pt+4.5*5pt,4.5*8.66pt);
         \filldraw [teal,fill=teal!60] (11*10pt+2*5pt,2*8.66pt) circle (2pt) node[right,black] {\scriptsize $\,=X$};
        \node at (0,4.35*8.66pt) {$\cdots$};
        \node at (130pt,4.35*8.66pt) {$\cdots$};
        \node at (6*10pt,5pt) {$\vdots$};
        \node at (-12pt+5*5pt,5*8.66pt) {\scriptsize $\partial \bar{A}$};
        \node at (-12pt+4*5pt,4*8.66pt) {\scriptsize $\partial A$};
    \end{tikzpicture}\,.
\end{align}
As $X_j\ket{\Omega} = \ket{\Omega}$, we have
\begin{equation}
    \lrangle{U_A^X} = \bra{\Omega^+}(U^{\rm{LG}})^{\dagger}U_A^XU^{\rm{LG}} \ket{\Omega^+}=\bra{\Omega^+}(U^X_A)^{\dagger}(U^{\rm{LG}})^{\dagger}U_A^XU^{\rm{LG}} \ket{\Omega^+}\,,
\end{equation}
To obtain $(U^X_A)^{\dagger}(U^{\rm{LG}})^{\dagger}U_A^XU^{\rm{LG}}$, we first calculate the conjugation of ${\rm{CC}}Z$, ${\rm{C}}Z$ and $Z$ gates by the global $\ZZ_2$ symmetry:
\begin{align}
    U^X Z_i (U^X)^{\dagger} &= -Z_{i} \label{eq:gate_conj_1}\\
    U^X U^{{\rm{C}}Z}_{jk}(U^X)^{\dagger} &= -Z_jZ_k U^{{\rm{C}}Z}_{jk}\label{eq:gate_conj_2}\\
    U^XU_{{\rm{CC}}Z}^{ijk}(U^X)^{\dagger} &= -Z_i Z_j Z_k U^{{\rm{C}}Z}_{ij}U^{{\rm{C}}Z}_{jk}U^{{\rm{C}}Z}_{ik}U^{{\rm{CC}}Z}_{ijk}\,.\label{eq:gate_conj_3}
\end{align}
from which we obtain the commutator for gates within $A$ as
\begin{align}
    (U_A^X)^{\dagger}(U^{\rm{LG}}_A)^{\dagger} U_A^X U_A^{\rm{\rm{LG}}} = (-1)^{V_A+E_A+F_A} \prod_{j\in \partial A} U^{{\rm{C}}Z}_{j,j+1}Z_{j}\,,
    \label{eq:sa_conj}
\end{align}
where $V_A$, $E_A$, and $F_A$ are the number of sites, edges and triangles in $A$. 
Note that $ (-1)^{V_A+E_A+F_A}=1$ due to Euler's formula.

However, for gates at the boundary of $A$ ($\partial A$), further treatment are needed.
To proceed, we shall decompose $U^{\rm{LG}}$ into $U^{\rm{LG}} = U^{\rm{LG}}_A U^{\rm{LG}}_{\bar{A}}U^{\rm{LG}}_{\partial A\partial \bar{A}}$, where
\begin{align}
    U^{\rm{LG}}_A &= \prod_{\bigtriangleup\in A} U_{ijk}^{{\rm{CC}}Z} \prod_{{\rm{edges}}\in A}U^{{\rm{C}}Z}_{ij} \prod_{j\in A}Z_j\\
    U^{\rm{LG}}_{\partial A\partial \bar{A}} &=  \prod_{\bigtriangleup\in \partial A\cup \partial \bar{A}} U_{ijk}^{{\rm{CC}}Z} \prod_{{\rm{edges}}\in \partial A\cup \partial \bar{A}}U^{{\rm{C}}Z}_{ij} \prod_{j\in \partial A\cup \partial \bar{A}}Z_j\,.
    \label{eq:ULG_papa}
\end{align}
Pictorially, the boundary degrees of freedom are 
\begin{equation}
    \partial A \cup \partial \bar{A}=\cdots\begin{tikzpicture}[scale=0.8,baseline={([yshift=-4.5pt]current bounding box.center)}]
        \def\unitcell{
        \draw[thick] (0pt , 0pt) --++ (30pt,0) --++(-15pt,15pt*1.732)--cycle;
        }
         \foreach \t in {0,1,2,3}
         {
            \begin{scope}[xshift=-15pt+\t*30pt]
            \unitcell;
         \end{scope}
         \draw node[below] at (\t*30pt-15pt,0){$s_{\t}$};
         }
         \draw node[below] at (4*30pt-15pt,0){$s_{4}$};
         \foreach \t in {0,1,2,3}
         {
            \begin{scope}[xshift=\t*30pt+15pt,rotate=60]
            \unitcell;
         \end{scope}
         \draw node[above] at (\t*30pt,15pt*1.73){$s_{\t}^\prime$};
         }
         \draw node[above] at (4*30pt,15pt*1.73){$s_{4}^\prime$};
         \draw[dashed] (-30pt,7.5pt*1.732)--+(160pt,0);
         \node at (-30pt,15pt*1.732) {$\partial\bar{A}$};
         \node at (-30pt,0) {$\partial{A}$};
    \end{tikzpicture}~\cdots\,,
\end{equation}
And $U^{\rm{LG}}_{\partial A\partial \bar{A}}$ in Eq.~\eqref{eq:ULG_papa} is expressed as
\begin{align}
    U^{\rm{LG}}_{\partial A\partial \bar{A}} = \prod_{j\in\partial A} U^{{\rm{CC}}Z}_{j,j^{\prime},j+1}U^{{\rm{CC}}Z}_{j^{\prime},j+1,(j+1)^{\prime}} U^{{\rm{C}}Z}_{j,j^{\prime}}U^{{\rm{C}}Z}_{j,j+1}U^{{\rm{C}}Z}_{j^{\prime},(j+1)^{\prime}}U^{{\rm{C}}Z}_{j^{\prime},j+1}Z_{j}Z_{j^{\prime}}\,.
\end{align}
From the following relations
\begin{align}
    X_j U^{{\rm{C}}Z}_{jk}X_j &= Z_{k}U^{{\rm{C}}Z}_{jk}\\
    X_i U^{{\rm{CC}}Z}_{ijk} X_i&= U^{{\rm{C}}Z}_{jk}U^{{\rm{CC}}Z}_{ijk}\\
    X_iX_j U^{{\rm{CC}}Z}_{ijk} X_iX_j&= Z_k U^{{\rm{C}}Z}_{ik}U^{{\rm{C}}Z}_{jk}U^{{\rm{CC}}Z}_{ijk}\,,
\end{align}
we obtain
\begin{align}
    (U_A^X)^{\dagger}(U^{\rm{LG}}_{\partial A\cup \partial \bar{A}})^{\dagger} U_A^X U_{\partial A\cup \partial \bar{A}}^{\rm{\rm{LG}}} = \prod_{j\in \partial A} Z_{j^{\prime}}U^{{\rm{C}}Z}_{j,j^{\prime}}U^{{\rm{C}}Z}_{j^{\prime}(j+1)^{\prime}}U^{{\rm{C}}Z}_{j^{\prime},j+1}\,,
    \label{eq:ba_conj}
\end{align}
Consequently, combining Eq.~\eqref{eq:ba_conj} and Eq.~\eqref{eq:sa_conj}, we have
\begin{equation}
    (U^X_A)^{\dagger}(U^{\rm{LG}})^{\dagger}U_A^XU^{\rm{LG}} 
    = \prod_{j\in \partial A} U^{{\rm{C}}Z}_{j,j+1}U^{{\rm{C}}Z}_{j,j^{\prime}}U^{{\rm{C}}Z}_{j^{\prime}(j+1)^{\prime}}U^{{\rm{C}}Z}_{j^{\prime},j+1}Z_{j}Z_{j^{\prime}}\,,
\end{equation}
which is an operator supported only on $\partial A\cup \partial \bar{A}$ as expected.

Therefore, $\bra{\Omega}(U^{\rm{LG}})^{\dagger}U_A^XU^{\rm{LG}}(U^X_A)^{\dagger}\ket{\Omega}$ can be evaluated by defining the transfer matrix 
\begin{equation}
    S^{s_j^{\prime}s_{j+1}^{\prime}}_{s_js_{j+1}} = U^{{\rm{C}}Z}_{j,j+1}U^{{\rm{C}}Z}_{j,j^{\prime}}U^{{\rm{C}}Z}_{j^{\prime}(j+1)^{\prime}}U^{{\rm{C}}Z}_{j^{\prime},j+1}Z_{j}Z_{j^{\prime}}\,,
\end{equation} 
which can be evaluated as
\begin{equation}
    S = I\otimes I + \ii I\otimes Y + \ii Y\otimes X + \ii Y\otimes Z\,,
\end{equation}
with $X$, $Y$, $Z$, and $I$ being Pauli matrices and identity.

We then conclude
\begin{equation}
    \lrangle{U_A^X} = \frac{1}{2^{2L_{\partial A}}}\sum_{s,s^{\prime}}\prod_jS^{s_j^{\prime}s_{j+1}^{\prime}}_{s_js_{j+1}} = \frac{1}{2^{2L_{\partial A}}}\Tr[(S)^{L_{\partial A}}] =\frac{4}{2^{L_{\partial A}}}\cos (\pi L_{\partial A}/3)\,,
\end{equation}
and hence
\begin{equation}
        -\ln \lrangle{U_A^X} = \ln 2\cdot L_{\partial A} - \ln (4\cos (\pi L_{\partial A}/3))\,.
\end{equation}

We now place two charged operator $Z_j$ and $Z_k$ at $\partial A$,
\begin{align}
     U_A^X Z_jZ_k=\begin{tikzpicture}[scale=0.9,baseline={([yshift=1pt]current bounding box.center)}]
        \usetikzlibrary{shapes.geometric}
        \foreach \x in {-2,-1,...,1}{
        \foreach \y in {2,3,...,7}{
           \filldraw[fill=black!50] (\x*10pt+\y*5pt,\y*8.66pt) circle (1pt);
        }
        }
         \foreach \x in {-2,-1,...,1}{
        \foreach \y in {2,3,4}{
           \filldraw[teal,fill=teal!60] (\x*10pt+\y*5pt,\y*8.66pt) circle (2pt);
        }
        }
        \foreach \x in {4,5,...,7}{
        \foreach \y in {2,3,...,7}{
           \filldraw[fill=black!50] (\x*10pt+\y*5pt,\y*8.66pt) circle (1pt);
        }
        }
         \foreach \x in {4,5,...,7}{
        \foreach \y in {2,3,4}{
           \filldraw[teal,fill=teal!60] (\x*10pt+\y*5pt,\y*8.66pt) circle (2pt);
        }
        }
         \node[diamond,draw=red,fill=red,scale=0.4] (d) at (0*10pt+4*5pt,4*8.66pt) {};
         \node[diamond,draw=red,fill=red,scale=0.4] (d) at (6*10pt+4*5pt,4*8.66pt) {};
         \filldraw [teal,fill=teal!60] (9*10pt+6*5pt,6*8.66pt) circle (2pt) node[right,black] {\scriptsize $\,=X$};
         \node[diamond,draw=red,fill=red,scale=0.4] (d) at (8*10pt+4*5pt,2*8.66pt){};
        \node[right,black]at (8*10pt+4*5pt,2*8.66pt) {\scriptsize $\,=XZ=-\ii Y$};
        \node at (-15pt,4*8.66pt) {$\cdots$};
        \node at (45pt,4*8.66pt) {$\cdots$};
         \node at (105pt,4*8.66pt) {$\cdots$};
         \node at (3.5*10pt,10pt) {$\vdots$};
    \end{tikzpicture}\,,
\end{align}
which gives
\begin{equation}
    \lrangle{ U_A^XZ_j Z_k}  = \bra{\Omega^+}(U_A^X)^{\dagger}(U^{\rm{LG}})^{\dagger}U_A^XU^{\rm{LG}}Z_jZ_k\ket{\Omega^+}\,,
\end{equation}
This quantity can also be calculated using the transfer matrix method, with additional two $Z\otimes I$ operators inserted:
\begin{align}
    &\lrangle{ U_A^X Z_j Z_k} = \frac{1}{2^{2L_{\partial A}}}\Tr[(S)^j\cdot Z\otimes I\cdot (S)^{k-j} \cdot Z\otimes I\cdot (S)^{L_{\partial A}-k}]\\
    =&\frac{4}{2^{L_{\partial A}}}\left[ \tan^2\frac{\pi}{6}\cos \frac{\pi L_{\partial A}}{3} + \left(1 - \tan^2\frac{\pi}{6}\right) \cos \frac{\pi (L_{\partial A}-2l)}{3}\right]
\end{align}
with $l=k-j$.
Finally, we obtain
\begin{equation}
    \langle Z_jZ_k \rangle_A^{g} = \frac{\lrangle{Z_jZ_kU_A^X}}{\lrangle{U_A^X}} = \frac{1 + 2\cos \frac{2\pi l }{3} +2\tan \frac{\pi L_{\partial A}}{3}\sin \frac{2\pi l}{3}}{3}\,.
\end{equation}

\section{Disorder operator of Levin-Gu state from RDM}
\label{sec:mpdo_lg}
In this part, we will evaluate the disorder operator of Levin-Gu state by solving the RDM in the entanglement space.
\subsection{PEPS representation of LG state}
We start by express LG state using PEPS.
For future reference, we express the amplitude $\braket{c}{\Psi_{\LG}}$ as a product of phase factors contributed from each triangle.
Consider the state $\ket{\alpha\beta\gamma}$ defined on the triangle $abc$, where $\alpha,\beta,\gamma\in\{0,1\}$.
The application of $U^{{\rm{CC}}Z}$ gates on this triangle give $(-1)^{\alpha\beta\gamma}$.
For a single edge, say edge $ab$, $U_{{\rm{C}}Z}$ gates give $(-1)^{\alpha\beta}$. 
Since each edge is shared by two adjacent triangles, the action of $U^{{\rm{C}}Z}$ gates on triangle $abc$ contributes $\exp \{\ii \frac{\pi}{2}(\alpha\beta+\beta\gamma+\alpha\gamma)\}$.
Regarding the action of the Pauli-${Z}$ gates on a single site, say $a$, introduces a phase factor $(-1)^\alpha$.
As each site is shared by six adjacent triangles, $Z$ gates give $\exp\{\ii\frac{\pi}{2}(\alpha+\beta+\gamma)\}$ to triangle $abc$.

Combining all these contributions of gates, it can be shown that the triangle $abc$ in state $\ket{\alpha\beta\gamma}$ gives an overall phase factor
\begin{equation}
    \xi_{abc}(\alpha,\beta,\gamma) = \exp\{\ii\frac{\pi}{2}(2\alpha\beta\gamma+\alpha\beta+\alpha\gamma+\beta\gamma+\alpha+\beta+\gamma) \}\,,
\end{equation}
And $\ket{\Psi^{\LG}}$ can then be rewritten as
\begin{equation}
    \ket{\Psi^{\LG}}=\frac{1}{2^{N/2}}\prod_{\bigtriangleup}\xi_{\bigtriangleup}\ket{c}
    \label{eq:lg_wf_triangle}
\end{equation}

Based on the preceding preparation, we are now ready to construct the PEPS representation of the LG state.
Let local tensor $\hat{T}$ accounts for the contribution from a unit cell comprising two triangles, depicted graphically as
\begin{align*}
    \hat{T}{}&=\begin{tikzpicture}[scale=1,baseline={([yshift=-2pt]current bounding box.center)}]
        \tikzstyle{wideline}=[cyan,line width=2pt]
        \def\unitcell{
            \draw[thick] (0pt , 0pt) --++ (30pt,0) --++(-15pt,15pt*1.732)--cycle;
        }
        \unitcell;
        \begin{scope}[xshift=30pt,rotate=60]
            \unitcell;
        \end{scope}
        \draw[wideline]  (10pt,-10pt)node[xshift=5pt,yshift=-5pt,black]{$d$} .. controls +(0,20pt) and +(20pt,0)..++(-20pt,20pt);
        \draw[wideline]  (20pt,-10pt) .. controls +(0,20pt) and +(-20pt,0)..++(20pt,20pt) --++(10pt,0);
        \draw[wideline]  (25pt,29.98pt+10pt)node[yshift=5pt,black]{$~~~u$} .. controls +(0,-20pt) and +(20pt,0)..++(-20pt,-20pt)--++(-15pt,0)node[xshift=-5pt,yshift=-5pt,black]{$l$};
        \draw[wideline]  (35pt,29.98pt+10pt) .. controls +(0,-20pt) and +(-20pt,0)..++(20pt,-20pt)node[xshift=5pt,yshift=-5pt,black]{$r$};
        \node at (0,0) {$\alpha$};
        \node at (30pt,0) {$\beta$};
        \node at (15pt,15pt*1.732) {$\gamma$};
        \node at (45pt,15pt*1.732) {$\delta$};
    \end{tikzpicture}=\sum_{p;u,l,d,r}T_p^{uldr}\ket{p}|u)|r)(d|(l|\\
    &=~~\begin{tikzpicture}[scale=1.5,baseline={([yshift=-2pt]current bounding box.center)}]
        \tikzstyle{wideline}=[cyan,line width=2pt]
        \tikzstyle ->-=[postaction={decorate,decoration={markings,
            mark=at position .55 with {\arrow[scale=0.7]{>}}}}]
            \tikzstyle -<-=[postaction={decorate,decoration={markings,
        mark=at position .35 with {\arrowreversed[scale=0.7]{>};}}}]
            \tikzstyle{wideline}=[cyan,line width=2pt]
        \filldraw[thick,fill=white] (0pt,-8pt) rectangle (16pt,8pt);
        \node at (8pt,0pt) {$T$};
        \draw[wideline,->-] (4pt,20pt) -- ++(0,-16pt) ;
        \draw[wideline,->-] (4pt,4pt) --++(-18pt,0);
        \draw[->-,thick]  (-6pt,14pt) -- (4pt,4pt);
        \node at (-8pt,16pt){$p$};
        \draw[wideline,->-] (12pt,20pt) -- ++(0,-16pt);
        \draw[wideline,-<-] (12pt,4pt)--++(18pt,0);
        \draw[wideline,-<-] (4pt,-20pt) -- ++(0,+16pt);
        \draw[wideline,->-] (4pt,-4pt)--++(-18pt,0);
        \draw[wideline,-<-] (12pt,-20pt) -- ++(0,+16pt);
        \draw[wideline,-<-] (12pt,-4pt) --++(18pt,0);
        \node at (-17pt,0pt){$l$};
        \node at (32pt,0pt){$r$};
        \node at (8pt,23pt){$u$};
        \node at (8pt,-23pt){$d$};
        \end{tikzpicture}
        =\begin{tikzpicture}[scale=1.5,baseline={([yshift=-2pt]current bounding box.center)}]
            \tikzstyle ->-=[postaction={decorate,decoration={markings,
            mark=at position .55 with {\arrow[scale=0.7]{>}}}}]
            \tikzstyle -<-=[postaction={decorate,decoration={markings,
        mark=at position .35 with {\arrowreversed[scale=0.7]{>};}}}]
            \tikzstyle{wideline}=[cyan,line width=2pt]
            \node at (-14pt,14pt){$p$};
            \node at (-23pt,0pt){$l$};
            \node at (23pt,0pt){$r$};
            \node at (0pt,23pt){$u$};
            \node at (0pt,-23pt){$d$};
            \draw[thick,->-] (0pt,20pt) --(0pt,0pt);
            \draw[thick,->-] (20pt,0pt) --(0pt,0pt);
            \draw[thick,-<-] (0,-20pt) --(0,0);
            \draw[thick,-<-] (-20pt,0) --(0,0);
            \draw[thick,-<-] (0,0)--++(-12pt,12pt);
        \end{tikzpicture}
\end{align*}
which hosts one physical index $p$ and four internal indices (virtual legs): $u$, $l$, $d$, and $r$.
Each virtual leg has a ``double-line" representation.
$\alpha$, $\beta$, $\gamma$, and $\delta$ denote Ising spins that take value in $\{0,1\}$.
The physical state of $\hat{T}$ is set to be identical to $\gamma$, i.e. $p\equiv\gamma$.
The blue lines copy the Ising spins within the enclosed area, and hence $\vket{u} = \vket{\delta\gamma}$, $\vbra{l} = \vbra{\gamma\alpha}$, $\vbra{d} = \vbra{\alpha\beta}$, and $\vket{r} = \vket{\beta\delta}$.
For the Ising spin configuration $\alpha\beta\gamma\delta$, the corresponding tensor entry is determined by $\xi$ of the two triangles:
\begin{equation}
    T_p^{uldr} \equiv F^{\alpha\beta\gamma\delta} 
    = \frac{\xi(\beta,\alpha,\gamma)\xi(\beta,\delta,\gamma)}{2}\,.
    \label{eq:F_entry}
\end{equation}
where we define $F$ for future usage.

To see the $\ZZ_2$ symmetry, we observe that tensor entry $F$ satisfies the following property
\begin{equation}
F^{\bar{\alpha}\bar{\beta}\bar{\gamma}\bar{\delta}} = F^{\alpha\beta\gamma\delta}(-1)^{(\gamma\delta+\gamma\alpha+\alpha\beta+\delta\beta + \alpha+\delta)},
\end{equation}
where $\bar{\alpha}=1-\alpha$, and similarly for $\bar{\beta}$, $\bar{\gamma}$, and $\bar{\delta}$.
This symmetry condition can be nicely organized into {\emph{tensor equations}} for $\hat{T}$:
\begin{equation}
    X_p W_u W_r \cdot \hat{T}\cdot W^{\dg}_d W^{\dg}_l  = \hat{T}\,,
\end{equation}
which is represented graphically as
\begin{equation}
\begin{tikzpicture}[scale=1.2,baseline={([yshift=0pt]current bounding box.center)}]
    \tikzstyle{wideline}=[cyan,line width=2pt]
    \filldraw[thick,fill=white] (0pt,-8pt) rectangle (16pt,8pt);
    \node at (8pt,0pt) {$\hat{T}$};
    \node[above,xshift=0pt] at (8pt,20pt){$W_u$};
    \node[below,xshift=0pt] at (8pt,-20pt){$W_d^{\dg}$};
    \node at (-10pt,18pt) {$X_p$};
    \draw[wideline] (4pt,20pt) -- ++(0,-16pt) --++(-18pt,0);
    \draw[thick] (4pt,4pt) -- ++(-10pt,10pt);
    \draw[wideline] (12pt,20pt) -- ++(0,-16pt) --++(18pt,0);
    \node[right] at (28pt,0pt){$W_r$};
    \node[left] at (-12pt,0pt){$W_l^{\dg}$};
    \draw[wideline] (4pt,-20pt) -- ++(0,+16pt) --++(-18pt,0);
    \draw[wideline] (12pt,-20pt) -- ++(0,+16pt) --++(18pt,0);
    \end{tikzpicture}=
    \begin{tikzpicture}[scale=1.2,baseline={([yshift=-2pt]current bounding box.center)}]
        \tikzstyle{wideline}=[cyan,line width=2pt]
        \filldraw[thick,fill=white] (0pt,-8pt) rectangle (16pt,8pt);
        \node at (8pt,0pt) {$\hat{T}$};
        \draw[thick] (4pt,4pt) -- ++(-10pt,10pt);
        \draw[wideline] (4pt,20pt) -- ++(0,-16pt) --++(-18pt,0);
        \draw[wideline] (12pt,20pt) -- ++(0,-16pt) --++(18pt,0);
        \draw[wideline] (4pt,-20pt) -- ++(0,+16pt) --++(-18pt,0);
        \draw[wideline] (12pt,-20pt) -- ++(0,+16pt) --++(18pt,0);
        \end{tikzpicture}\,,
        \label{eq:sym_tensor_z2spt_peps}
\end{equation}
Here,
\begin{equation}
    \begin{aligned}
        X_p \ket{p} &= \ket{\bar{p}}~;\quad
        W_u\vket{\gamma \delta} = (-1)^{\gamma\delta}(\ii)^{\delta+\gamma}\vbra{\bar{\gamma},\bar{\delta}}~,\quad
        \vbra{\gamma \alpha}W^{\dg}_l = (-1)^{\gamma\alpha}(-\ii)^{\alpha+\gamma}\vbra{\bar{\gamma},\bar{\alpha}}~,\\
        \vbra{\alpha \beta}W^{\dg}_d &= (-1)^{\alpha\beta}(-\ii)^{\alpha+\beta}\vbra{\bar{\alpha},\bar{\beta}}~,\quad
        W_r\vket{\delta \beta} = (-1)^{\delta\beta}(\ii)^{\delta+\beta}\vket{\bar{\delta},\bar{\beta}}\,.
    \end{aligned}
\end{equation}
In terms of matrix form, we have
\begin{equation}
    \begin{aligned}
        W &=\vket{00}\vbra{11}+\ii\vket{10}\vbra{01}+\ii\vket{01}\vbra{10}+\vket{11}\vbra{00}\\
        &= X\otimes X\exp\left\{\ii\frac{\pi}{4}(1-Z\otimes Z)\right\}\,,
    \end{aligned}
    \label{eq:w_def}
\end{equation}
Therefore, symmetry action on physical legs are passed to internal legs as a gauge transformation, resulting in a symmetric wavefunction.

We point out that symmetry action on internal legs here form a $\ZZ_4$ group, where
\begin{equation}
    W^2\vket{\alpha\beta} = (-1)^{\alpha+\beta}\vket{\alpha\beta} =Z_\alpha Z_{\beta} |\alpha\beta)\,,
    \label{eq:w2d2z}
\end{equation} 
Note that $Z$ action on legs within a plaquette leaves tensor invariant, which form a plaquette invariant gauge group (IGG)\cite{spt_jiang2016}:
\begin{equation}
\begin{tikzpicture}[scale=1.2,baseline={([yshift=-8pt]current bounding box.center)}]
    \tikzstyle{wideline}=[cyan,line width=2pt]
    \filldraw[thick,fill=white] (0pt,-8pt) rectangle (16pt,8pt);
    \node at (8pt,0pt) {$\hat{T}$};
    \node[above,xshift=0pt] at (12pt,20pt){$Z$};
    \draw[wideline] (4pt,20pt) -- ++(0,-16pt) --++(-18pt,0);
    \draw[thick] (4pt,4pt) -- ++(-10pt,10pt);
    \draw[wideline] (12pt,20pt) -- ++(0,-16pt) --++(18pt,0);
    \node[right] at (28pt,4pt){$Z$};
    \draw[wideline] (4pt,-20pt) -- ++(0,+16pt) --++(-18pt,0);
    \draw[wideline] (12pt,-20pt) -- ++(0,+16pt) --++(18pt,0);
    \end{tikzpicture}=
    \begin{tikzpicture}[scale=1.2,baseline={([yshift=-2pt]current bounding box.center)}]
        \tikzstyle{wideline}=[cyan,line width=2pt]
        \filldraw[thick,fill=white] (0pt,-8pt) rectangle (16pt,8pt);
        \node at (8pt,0pt) {$\hat{T}$};
        \draw[thick] (4pt,4pt) -- ++(-10pt,10pt);
        \draw[wideline] (4pt,20pt) -- ++(0,-16pt) --++(-18pt,0);
        \draw[wideline] (12pt,20pt) -- ++(0,-16pt) --++(18pt,0);
        \draw[wideline] (4pt,-20pt) -- ++(0,+16pt) --++(-18pt,0);
        \draw[wideline] (12pt,-20pt) -- ++(0,+16pt) --++(18pt,0);
        \end{tikzpicture}\,
        \label{eq:plq_igg_z2spt_peps}
\end{equation}
Finally, the last important piece is the anti-commutation relation 
\begin{equation}
WZ W^{\dg} = -Z
\label{eq:anti_comm_sym_plq_igg}
\end{equation}
We will explain in detail in Sec.~\ref{sec:rdm_spt} about how these tensor equations -- including Eq.~\eqref{eq:sym_tensor_z2spt_peps}, Eq.~\eqref{eq:plq_igg_z2spt_peps}, Eq.~\eqref{eq:w2d2z}, and Eq.~\eqref{eq:anti_comm_sym_plq_igg} -- lead to the $\ZZ_2$ SPT phase.

\subsection{MPDO for LG state}
Equipped with the PEPS representation of LG state, the entanglement space $\mathcal{H}_{\partial A}$ between lower half plane $A$ and its complement $\bar{A}$ is identified as virtual legs between them, i.e. $\mathcal{H}_{\partial A} = \otimes_{j=1}^{L_{\partial A}}\mathcal{H}_j$, where $\mathcal{H}_j$ labels virtual leg $j$.
When symmetry is acted on $A$, it is passed to $\HH_{\partial A}$ as illustrated below
\begin{align}
    \begin{tikzpicture}[scale=1,baseline={([yshift=0pt]current bounding box.center)}]
    \definecolor{myorange}{rgb}{1.0,0.5,0}
    \draw[thick] (-15pt,0pt) --(15pt,0pt);
    \draw[thick] (0,-15pt) --(0,15pt);
    \draw[thick] (0,0)--++(-10pt,8pt);
    \filldraw[black,fill=black] (-5pt,4pt) circle (1pt) node[xshift=-5pt,yshift=8pt] {\scriptsize $U(g)$};
    \node at (0,-20pt){$=$};
    \begin{scope}[yshift=-40pt]
        \draw[thick] (-15pt,0pt) --(15pt,0pt);
        \draw[thick] (0,-15pt) --(0,15pt);
        \draw[thick] (0,0)--++(-10pt,8pt);
        \filldraw[myorange,fill=myorange] (0,10pt) circle (1pt) node[above left] {\scriptsize $W$};
        \filldraw[myorange,fill=myorange] (-10pt,0) circle (1pt) node[xshift=-10pt] {\scriptsize $W^{\dg}$};
        \filldraw[myorange,fill=myorange] (0,-10pt) circle (1pt) node[below left] {\scriptsize $W^{\dg}$};
        \filldraw[myorange,fill=myorange] (10pt,0) circle (1pt) node[right] {\scriptsize $W$};
    \end{scope}
    \end{tikzpicture} \Rightarrow
    \begin{tikzpicture}[scale=1,baseline={([yshift=-4pt]current bounding box.center)}]
    \definecolor{myorange}{rgb}{1.0,0.5,0}
    \def\myL{20pt};
    \foreach \t in {0,1,2,3}{
            \draw[thick] (-0.5*\myL,\t*\myL)--(3.5*\myL,\t*\myL); 
            \draw[thick] (\t*\myL,-0.5*\myL) -- (\t*\myL,3.5*\myL); 
        }
        \foreach \x in {0,1,2,3}{
        \foreach \y in {0,1,2,3}{
            \draw[thick](\x*\myL,\y*\myL)--++(-8pt,8pt);
        }
        }
        \foreach \x in {0,1,2,3}{
        \foreach \y in {0,1}{
            \draw[thick](\x*\myL,\y*\myL)--++(-8pt,8pt);
            \filldraw[black,fill=black] (\x*\myL-4pt,\y*\myL+4pt) circle (1pt) node[xshift=-5pt,yshift=8pt] {\scriptsize $U(g)$};
        }
        }
        \foreach \t in {0,1,2,3} {
         \node[right] at (3.5*\myL,\t*\myL) {\scriptsize$\cdots$};
         \node[left] at (-0.5*\myL,\t*\myL) {\scriptsize$\cdots$};
         \node[below] at (\t*\myL,-0.5*\myL) {\scriptsize$\vdots$};
         \node[above] at (\t*\myL,3.5*\myL) {\scriptsize$\vdots$};
        }
        \node at (90pt,30pt){$=$};
    \begin{scope}[xshift=120pt]
    \def\myL{20pt};
    \foreach \t in {0,1,2,3}
        {\draw[thick] (-0.5*\myL,\t*\myL)--(3.5*\myL,\t*\myL); 
        \draw[thick] (\t*\myL,-0.5*\myL) -- (\t*\myL,3.5*\myL); 
        }
        \foreach \x in {0,1,2,3}{
        \foreach \y in {0,1,2,3}{
            \draw[thick](\x*\myL,\y*\myL)--++(-8pt,8pt);
        }
        } 
        \foreach \t in {0,1,2,3} {
            \node[right] at (3.5*\myL,\t*\myL) {\scriptsize$\cdots$};
            \node[left] at (-0.5*\myL,\t*\myL) {\scriptsize$\cdots$};
            \node[below] at (\t*\myL,-0.5*\myL) {\scriptsize$\vdots$};
            \node[above] at (\t*\myL,3.5*\myL) {\scriptsize$\vdots$};
        }
        \foreach \t in {0,1,2,3}{
            \filldraw[myorange,fill=myorange] (\t*\myL,1.5*\myL) circle (1pt) node[xshift=-10pt,yshift=5pt] {\scriptsize $W(g)$};
        }
    \end{scope}
    \end{tikzpicture}   
    \label{eq:disorder_pass_virtual}
\end{align}
We now solve the RDM $\rho_{\partial A}$ through this tensor representation.
A bipartition of wavefunction is straightforward in PEPS:
\begin{align}
    \ket{\psi^{\rm{LG}}} = \Tr_{\mathcal{H}_{\partial A}}[ \hat{T}_A \otimes \hat{T}_{\bar{A}}]\,,
\end{align}
where
\begin{align}
    \hat{T}_A = \Tr_{\rm{virtual~space}}[\otimes_{j\in A} \hat{T}_j];~\hat{T}_{\bar{A}} = \Tr_{\rm{virtual~space}}[\otimes_{j\in {\bar{A}}} \hat{T}_j]\,.
\end{align}
The RDM $\rho_A$ is given by
\begin{align}
    \rho_A =\Tr_{\bar{A}}\ket{\psi^{\rm{LG}}}\bra{\psi^{\rm{LG}}} = \Tr_{\bar{A},\mathcal{H}_{\partial A}} [\hat{T}_A \otimes \hat{T}_{\bar{A}}\otimes \hat{T}_{\bar{A}}^{\dagger}\otimes \hat{T}_A^{\dagger}]\,.
\end{align}
It is easy to check that
\begin{align}
    \Tr_{A}[\hat{T}_{A}^{\dagger}\otimes\hat{T}_{A} ] = \mathbf{1}_{\mathcal{H}_{\partial A}}\,,
\end{align}
Namely, $\hat{T}_A$ is the isometry map from $\mathcal{H}_A$ to $\mathcal{H}_{\partial A}$.
Consequently, we shall define the density operator on $\HH_{\partial A}$ as
\begin{align}
    \rho_{\partial A} = \Tr_{\bar{A}} [\hat{T}_{\bar{A}}\otimes \hat{T}_{\bar{A}}^{\dagger}]
\end{align}

By summing over all Ising spins in $\bar{A}$, one obtains
\begin{align*}
    \Tr_{\bar{A}} [\hat{T}_{\bar{A}}\otimes \hat{T}_{\bar{A}}^{\dagger}]=\cdots\begin{tikzpicture}[scale=1.2,baseline={([yshift=-2pt]current bounding box.center)}]
        \tikzstyle{wideline}=[cyan,line width=2pt]
        \filldraw[thick,fill=white] (0pt,-8pt) rectangle (16pt,8pt);
        \node at (8pt,0pt) {$\hat{T}$};
        \node at (4pt,-24pt) {$\scriptsize\alpha_0$};
        \node at (12pt,-24pt) {$\scriptsize\alpha_1$};
        \draw[thick] (-8pt,20pt) arc (180:232:10pt);
        \draw[thick] (4pt,4pt) -- ++(-10pt,10pt);
        \draw[wideline] (4pt,20pt) -- ++(0,-16pt) --++(-18pt,0);
        \draw[wideline] (12pt,20pt) -- ++(0,-16pt) --++(18pt,0);
        \draw[wideline] (4pt,-20pt) -- ++(0,+16pt) --++(-18pt,0);
        \draw[wideline] (12pt,-20pt) -- ++(0,+16pt) --++(18pt,0);
        \begin{scope}[xshift=40pt]
            \filldraw[thick,fill=white] (0pt,-8pt) rectangle (16pt,8pt);
            \node at (8pt,0pt) {$\hat{T}$};
        \node at (4pt,-24pt) {$\scriptsize\alpha_1$};
        \node at (12pt,-24pt) {$\scriptsize\alpha_2$};
        \draw[thick] (-8pt,20pt) arc (180:232:10pt);
            \draw[thick] (4pt,4pt) -- ++(-10pt,10pt);
            \draw[wideline] (4pt,20pt) -- ++(0,-16pt) --++(-18pt,0);
            \draw[wideline] (12pt,20pt) -- ++(0,-16pt) --++(18pt,0);
            \draw[wideline] (4pt,-20pt) -- ++(0,+16pt) --++(-18pt,0);
            \draw[wideline] (12pt,-20pt) -- ++(0,+16pt) --++(18pt,0);
        \end{scope}
        \begin{scope}[xshift=80pt]
            \filldraw[thick,fill=white] (0pt,-8pt) rectangle (16pt,8pt);
            \node at (8pt,0pt) {$\hat{T}$};
        \node at (4pt,-24pt) {$\scriptsize\alpha_2$};
        \node at (12pt,-24pt) {$\scriptsize\alpha_3$};
        \draw[thick] (-8pt,20pt) arc (180:232:10pt);
            \draw[thick] (4pt,4pt) -- ++(-10pt,10pt);
            \draw[wideline] (4pt,20pt) -- ++(0,-16pt) --++(-18pt,0);
            \draw[wideline] (12pt,20pt) -- ++(0,-16pt) --++(18pt,0);
            \draw[wideline] (4pt,-20pt) -- ++(0,+16pt) --++(-18pt,0);
            \draw[wideline] (12pt,-20pt) -- ++(0,+16pt) --++(18pt,0);
        \end{scope}
        \begin{scope}[xshift=120pt]
            \filldraw[thick,fill=white] (0pt,-8pt) rectangle (16pt,8pt);
            \node at (8pt,0pt) {$\hat{T}$};
        \node at (4pt,-24pt) {$\scriptsize\alpha_3$};
        \node at (12pt,-24pt) {$\scriptsize\alpha_4$};
        \draw[thick] (-8pt,20pt) arc (180:232:10pt);
            \draw[thick] (4pt,4pt) -- ++(-10pt,10pt);
            \draw[wideline] (4pt,20pt) -- ++(0,-16pt) --++(-18pt,0);
            \draw[wideline] (12pt,20pt) -- ++(0,-16pt) --++(18pt,0);
            \draw[wideline] (4pt,-20pt) -- ++(0,+16pt) --++(-18pt,0);
            \draw[wideline] (12pt,-20pt) -- ++(0,+16pt) --++(18pt,0);
        \end{scope}
        \begin{scope}[yshift=40pt]
            \filldraw[thick,fill=white] (0pt,-8pt) rectangle (16pt,8pt);
            \node at (8pt,0pt) {$\,\,\hat{T}^{\dagger}$};
            \node at (4pt,24pt) {$\scriptsize\beta_0$};
            \node at (12pt,24pt) {$\scriptsize\beta_1$};
            \draw[thick] (-8pt,-20pt) arc (180:128:10pt);
                \draw[thick] (4pt,-4pt) -- ++(-10pt,-10pt);
            \draw[wideline] (4pt,20pt) -- ++(0,-16pt) --++(-18pt,0);
            \draw[wideline] (12pt,20pt) -- ++(0,-16pt) --++(18pt,0);
            \draw[wideline] (4pt,-20pt) -- ++(0,+16pt) --++(-18pt,0);
            \draw[wideline] (12pt,-20pt) -- ++(0,+16pt) --++(18pt,0);
            \begin{scope}[xshift=40pt]
                \filldraw[thick,fill=white] (0pt,-8pt) rectangle (16pt,8pt);
                \node at (8pt,0pt) {$\,\,\hat{T}^{\dagger}$};
            \node at (4pt,24pt) {$\scriptsize\beta_1$};
            \node at (12pt,24pt) {$\scriptsize\beta_2$};
            \draw[thick] (-8pt,-20pt) arc (180:128:10pt);
                \draw[thick] (4pt,-4pt) -- ++(-10pt,-10pt);
                \draw[wideline] (4pt,20pt) -- ++(0,-16pt) --++(-18pt,0);
                \draw[wideline] (12pt,20pt) -- ++(0,-16pt) --++(18pt,0);
                \draw[wideline] (4pt,-20pt) -- ++(0,+16pt) --++(-18pt,0);
                \draw[wideline] (12pt,-20pt) -- ++(0,+16pt) --++(18pt,0);
            \end{scope}
            \begin{scope}[xshift=80pt]
                \filldraw[thick,fill=white] (0pt,-8pt) rectangle (16pt,8pt);
                \node at (8pt,0pt) {$\,\,\hat{T}^{\dagger}$};
            \node at (4pt,24pt) {$\scriptsize\beta_2$};
            \node at (12pt,24pt) {$\scriptsize\beta_3$};
            \draw[thick] (-8pt,-20pt) arc (180:128:10pt);
                \draw[thick] (4pt,-4pt) -- ++(-10pt,-10pt);
                \draw[wideline] (4pt,20pt) -- ++(0,-16pt) --++(-18pt,0);
                \draw[wideline] (12pt,20pt) -- ++(0,-16pt) --++(18pt,0);
                \draw[wideline] (4pt,-20pt) -- ++(0,+16pt) --++(-18pt,0);
                \draw[wideline] (12pt,-20pt) -- ++(0,+16pt) --++(18pt,0);
            \end{scope}
            \begin{scope}[xshift=120pt]
                \filldraw[thick,fill=white] (0pt,-8pt) rectangle (16pt,8pt);
                \node at (8pt,0pt) {$\,\,\hat{T}^{\dagger}$};
            \node at (4pt,24pt) {$\scriptsize\beta_3$};
            \node at (12pt,24pt) {$\scriptsize\beta_4$};
            \draw[thick] (-8pt,-20pt) arc (180:128:10pt);
                \draw[thick] (4pt,-4pt) -- ++(-10pt,-10pt);
                \draw[wideline] (4pt,20pt) -- ++(0,-16pt) --++(-18pt,0);
                \draw[wideline] (12pt,20pt) -- ++(0,-16pt) --++(18pt,0);
                \draw[wideline] (4pt,-20pt) -- ++(0,+16pt) --++(-18pt,0);
                \draw[wideline] (12pt,-20pt) -- ++(0,+16pt) --++(18pt,0);
            \end{scope}
        \end{scope} 
        \end{tikzpicture}\cdots\,,
\end{align*}
with
\begin{align*}
    \begin{tikzpicture}[scale=1.2,baseline={([yshift=-2pt]current bounding box.center)}]
        \tikzstyle{wideline}=[cyan,line width=2pt]
        \filldraw[thick,fill=white] (0pt,-8pt) rectangle (16pt,8pt);
                \node at (8pt,0pt) {$\,\,\hat{T}^{\dagger}$};
            \node at (-10pt,-18pt) {$\scriptsize\sigma_j$};
            \node at (4pt,-24pt) {$\scriptsize\sigma_j$};
            \node at (4pt,24pt) {$\scriptsize\beta_j$};
            \node at (16pt,-24pt) {$\scriptsize\sigma_{j+1}$};
            \node at (16pt,24pt) {$\scriptsize\beta_{j+1}$};
                \draw[thick] (4pt,-4pt) -- ++(-10pt,-10pt);
                \draw[wideline] (4pt,20pt) -- ++(0,-16pt) --++(-18pt,0);
                \draw[wideline] (12pt,20pt) -- ++(0,-16pt) --++(18pt,0);
                \draw[wideline] (4pt,-20pt) -- ++(0,+16pt) --++(-18pt,0);
                \draw[wideline] (12pt,-20pt) -- ++(0,+16pt) --++(18pt,0);
    \end{tikzpicture}=(F^{\beta_j \beta_{j+1}\sigma_j \sigma_{j+1}})^*\,.
\end{align*}

Therefore, $\rho_{\partial A}$ is naturally an MPDO
\begin{equation}
    \rho_{\partial A} = \Tr \left[M^{L_{\partial A}} \right]
    ~=~ \begin{tikzpicture}[scale=1,baseline={([yshift=-2pt]current bounding box.center)}]
        \draw[white] (0pt,-20pt) -- (0pt,20pt);
        \draw (-8pt,0pt) -- (46pt,0pt);
        \draw[densely dashed] (46pt,0pt) -- (78pt,0pt);
        \draw (78pt,0pt) -- (104pt,0pt);
        \draw (-8pt,0pt) arc (90:270:2.5pt);
        \draw (104pt,0pt) arc (90:-90:2.5pt);
        \draw [densely dotted] (-8pt,-5pt) -- (-2pt,-5pt);
        \draw [densely dotted] (104pt,-5pt) -- (98pt,-5pt);
        \draw (8pt,-16pt) -- (8pt,16pt);
        \draw (36pt,-16pt) -- (36pt,16pt);
        \draw (88pt,-16pt) -- (88pt,16pt);
        \filldraw[thick,fill=white] (0pt,-8pt) rectangle (16pt,8pt);
        \node at (8pt,0pt) {$M$};
        \filldraw[thick,fill=white] (28pt,-8pt) rectangle (44pt,8pt);
        \node at (36pt,0pt) {$M$};
        \filldraw[thick,fill=white] (80pt,-8pt) rectangle (96pt,8pt);
        \node at (88pt,0pt) {$M$};
    \end{tikzpicture}
\end{equation}
in which
\begin{align}
 \begin{tikzpicture}[scale=1,baseline={([yshift=-2pt]current bounding box.center)}]
        \draw (8pt,-20pt) -- (8pt,20pt);
        \draw (-12pt,0pt) -- (28pt,0pt);
        \filldraw[thick,fill=white] (0pt,-8pt) rectangle (16pt,8pt);
        \node at (8pt,0pt) {$M$};
        \end{tikzpicture}
        =
        \begin{tikzpicture}[scale=1.2,baseline={([yshift=-2pt]current bounding box.center)}]
            \tikzstyle{wideline}=[cyan,line width=2pt]
            \draw (-12pt,0pt) -- (28pt,0pt);
            \node[left] at (-12pt,0){\scriptsize$\sigma_\ell$};
            \node[right] at (28pt,0){\scriptsize$\sigma_r$};
            \filldraw[thick,fill=white] (0pt,-8pt) rectangle (16pt,8pt);
            \node at (8pt,0pt) {\scriptsize$M$};
            \node[above,xshift=-4pt] at (6pt,20pt){\scriptsize$\beta_\ell$};
            \draw[wideline] (6pt,20pt) -- ++(0,-14pt) --++(-18pt,0);
            \node[left] at (-12pt,6pt){\scriptsize$\beta_\ell$};
            \node[above,xshift=4pt] at (10pt,20pt){\scriptsize$\beta_r$};
            \draw[wideline] (10pt,20pt) -- ++(0,-14pt) --++(18pt,0);
            \node[right] at (28pt,6pt){\scriptsize $\beta_r$};
            \node[below,xshift=-4pt] at (6pt,-20pt){\scriptsize $\alpha_\ell$};
            \draw[wideline] (6pt,-20pt) -- ++(0,+14pt) --++(-18pt,0);
            \node[below,xshift=4pt] at (10pt,-20pt){\scriptsize $\alpha_r$};
            \node[left] at (-12pt,-6pt){\scriptsize$\alpha_\ell$};
            \draw[wideline] (10pt,-20pt) -- ++(0,+14pt) --++(18pt,0);
            \node[right] at (28pt,-6pt){\scriptsize $\alpha_r$};
            \end{tikzpicture}
        = \begin{aligned}[]
            \left(F^{\beta_\ell\beta_r\sigma_\ell\sigma_r}\right)^* \cdot F^{\alpha_\ell\alpha_r\sigma_\ell\sigma_r}
        \end{aligned}
\end{align}
with $F$ defined in Eq.~\eqref{eq:F_entry}. 

With $\rho_{\partial A}$, the disorder parameter $\lrangle{U_A(g)}$ is expressed as
\begin{align}
    \Tr_A[U_A(g)\rho_A] = \Tr_{A} [U_A(g)\cdot\hat{T}_A\cdot \rho_{\partial A}\cdot \hat{T}_A^{\dagger}]=\Tr_{\mathcal{H}_{\partial A}} [ \hat{T}_A^{\dagger}\cdot U_A(g)\cdot\hat{T}_A\cdot \rho_{\partial A}]=\Tr_{\mathcal{H}_{\partial A}}[W_{\partial A} \rho_{\partial A}]\,,
\end{align}
with $W$ defined in Eq.~\eqref{eq:w_def}.
In terms of MPDO, we have
\begin{equation}
    \begin{aligned}
        \Tr [W_{\partial A} \rho_{\partial A}]
        =  
        \begin{tikzpicture}[scale=1,baseline={([yshift=-2pt]current bounding box.center)}]
            \draw (8pt,-20pt) -- (8pt,20pt);
            \draw (8pt,20pt) arc (180:0:2.5pt);
            \draw (8pt,-20pt) arc (180:360:2.5pt);
            \draw [densely dotted] (13pt,20pt) -- (13pt,8pt);
            \draw [densely dotted] (13pt,-20pt) -- (13pt,-8pt);
            \draw (36pt,-20pt) -- (36pt,20pt);
            \draw (36pt,20pt) arc (180:0:2.5pt);
            \draw (36pt,-20pt) arc (180:360:2.5pt);
            \draw [densely dotted] (41pt,20pt) -- (41pt,8pt);
            \draw [densely dotted] (41pt,-20pt) -- (41pt,-8pt);
            \draw (88pt,-20pt) -- (88pt,20pt);
            \draw (88pt,20pt) arc (180:0:2.5pt);
            \draw (88pt,-20pt) arc (180:360:2.5pt);
            \draw [densely dotted] (93pt,20pt) -- (93pt,8pt);
            \draw [densely dotted] (93pt,-20pt) -- (93pt,-8pt);
            \draw[white] (0pt,-20pt) -- (0pt,20pt);
            \draw (-8pt,0pt) -- (46pt,0pt);
            \draw[densely dashed] (46pt,0pt) -- (78pt,0pt);
            \draw (78pt,0pt) -- (104pt,0pt);
            \draw (-8pt,0pt) arc (90:270:2.5pt);
            \draw (104pt,0pt) arc (90:-90:2.5pt);
            \draw [densely dotted] (-8pt,-5pt) -- (-2pt,-5pt);
            \draw [densely dotted] (104pt,-5pt) -- (98pt,-5pt);
            \draw (8pt,-16pt) -- (8pt,16pt);
            \draw (36pt,-16pt) -- (36pt,16pt);
            \draw (88pt,-16pt) -- (88pt,16pt);
            \filldraw[thick,fill=white] (0pt,-8pt) rectangle (16pt,8pt);
            \node at (8pt,0pt) {$M$};
            \filldraw[thick,fill=white] (28pt,-8pt) rectangle (44pt,8pt);
            \node at (36pt,0pt) {$M$};
            \filldraw[thick,fill=white] (80pt,-8pt) rectangle (96pt,8pt);
            \node at (88pt,0pt) {$M$};
              \filldraw [red,fill=red] (88pt,15pt) circle (1pt) node[left] {\scriptsize $W(g)$};
                \filldraw [red,fill=red] (8pt,15pt) circle (1pt) node[left] {\scriptsize $W(g)$};
                 \filldraw [red,fill=red] (36pt,15pt) circle (1pt) node[left]{\scriptsize $W(g)$};
            \end{tikzpicture}
            = \Tr ((T(g))^L)
    \end{aligned}
\end{equation}
$T(g)$ can be calculated straightforwardly, whose non-zero entries are
\begin{align*}
    \begin{tikzpicture}[scale=1.2,baseline={([yshift=-8pt]current bounding box.center)}]
        \tikzstyle{wideline}=[cyan,line width=2pt]
        \coordinate (upleft) at (6pt,30pt);
        \coordinate (upright) at (10pt,30pt);
        \coordinate (downleft) at (6pt,-20pt);
        \coordinate (downright) at (10pt,-20pt);
        \draw (-12pt,0pt) -- (28pt,0pt);
        \node[left] at (-12pt,0){\scriptsize$\sigma_\ell$};
        \node[right] at (28pt,0){\scriptsize$\sigma_r$};
        \filldraw[thick,fill=white] (0pt,-8pt) rectangle (16pt,8pt);
        \node at (8pt,0pt) {\scriptsize$M$};
        \draw[wideline] (upleft) arc (0:180:2pt);
        \draw[wideline] (upleft) -- ++(0,-24pt) --++(-18pt,0);
        \node[left] at (-12pt,6pt){\scriptsize$\alpha_\ell$};
        \draw[wideline] (upright) arc (180:0:2pt);
        \draw[wideline] (upright) -- ++(0,-24pt) --++(18pt,0);
        \node[right] at (28pt,6pt){\scriptsize $\alpha_r$};
        \draw[wideline] (downleft) arc (0:-180:2pt);
        \draw [wideline,densely dotted] (2pt,-20pt) -- ++(0,10pt);
        \draw[wideline] (downleft) -- ++(0,+14pt) --++(-18pt,0);
        \node[left] at (-12pt,-6pt){\scriptsize$\bar{\alpha}_\ell$};
        \draw[wideline] (downright) arc (-180:0:2pt);
        \draw [wideline,densely dotted] (14pt,-20pt) -- ++(0,10pt);
        \draw[wideline] (10pt,-20pt) -- ++(0,+14pt) --++(18pt,0);
        \node[right] at (28pt,-6pt){\scriptsize $\bar{\alpha}_r$};
        \node at (8pt,20pt) [fill=white]{\scriptsize$W(g)$};
        \draw[wideline,densely dotted] (2pt,30pt) -- ++(0,-15pt);
        \draw[wideline,densely dotted] (14pt,30pt) -- ++(0,-15pt);
        \node at (8pt,20pt) [fill=white]{\scriptsize$W(g)$};
        \end{tikzpicture}
        = \frac{1}{4}\left(I_{\alpha_l\alpha_r}  I_{\sigma_l\sigma_r} - \ii Z_{\alpha_l\alpha_r} X_{\sigma_l\sigma_r} - \ii Y_{\alpha_l\alpha_r} I_{\sigma_l\sigma_r} - \ii X_{\alpha_l\alpha_r}  X_{\sigma_l\sigma_r}\right)\,,
\end{align*}
with $\bar{\alpha} = 1-\alpha$.
$T(g)$ hosts unitary $\ZZ_2$ symmetry generated by $V(g)=\ii YY$, as well as anti-unitary symmetries $\mathcal{H} = \ii YI\mathcal{K}$ and $\mathcal{T} =  IZ\mathcal{K}$, with $\mathcal{K}$ being the complex conjugation. 
Note that $\mathcal{H}^2=-1$, leading to Kramers doublets.

Spectrum of $T(g)$ are 4-fold degenerate in modulus.
We label the four eigenstates as as $\ket{\psi^n_v}$ with $n,v=\pm$, with eigenvalues $m_v^n=\exp(\ii n \pi/3)/2$.
Here, $V(g)\ket{\psi^n_v} = \ii v \ket{\psi^n_v}$, and these states are connected through $\TT$ and $\HH$:
\begin{align*}
    \begin{tikzpicture}
        \node[below] (1) at (0,0){$\ket{\psi_{-}^{-}}$};
        \node[below] (2) at (2,0){$\ket{\psi_{-}^{+}}$};
        \node[below] (3) at (0,2){$\ket{\psi_{+}^{+}}$};
        \node[below] (4) at (2,2){$\ket{\psi_{+}^{-}}$};
        \draw[<->,>=stealth] (1) to node[above]{$\mathcal{H}$} (2) ;
        \draw[<->,>=stealth] (1) to node[left]{$\mathcal{T}$} (3) ;
        \draw[<->,>=stealth] (4) to node[above]{$\mathcal{H}$} (3) ;
        \draw[<->,>=stealth] (4) to node[right]{$\mathcal{T}$} (2) ;
    \end{tikzpicture}
\end{align*}

With the above information, $\lrangle{U_A(g)}$ can be solved as 
\begin{align}
    \lrangle{U_A(g)} = \sum_{n;\,v} (m^n_v)^L = \frac{4}{2^L}\cos (\pi L/3)\,.
    \label{eq:disorder_op_lg_transfer_mat}
\end{align}

The expectation value of $\lrangle{Z_j Z_{j+l}U_A(g)}$ for the LG state can be calculated as
\begin{align}
    \lrangle{Z_j Z_{j+l}U_A(g)} 
    = \Tr [Z_\alpha T(g)^l Z_\alpha T(g)^{L-l}]\,.
\end{align}
As $Z_\alpha$ is odd under $V = \ii Y_\alpha Y_\sigma$, the above equation becomes 
\begin{equation}
  \sum_{v,n,n^{\prime}}|\bra{\psi^{n^{\prime}}_{-v}}Z_\alpha \ket{\psi^n_v}|^2 (m^{n^{\prime}}_{-v})^l (m^n_v)^{L-l}
    = \frac{4}{2^L}\left[ \tan^2\frac{\pi}{6}\cos \frac{\pi L}{3} + \left(1 - \tan^2\frac{\pi}{6}\right) \cos \frac{\pi (L-2l)}{3}\right]
    \label{eq:imo_lg_transfer_mat}
\end{equation}
Combining Eq.~\eqref{eq:disorder_op_lg_transfer_mat} and Eq.~\eqref{eq:imo_lg_transfer_mat}, we conclude
\begin{align}
   \lrangle{Z_j Z_{j+l}}_{A}^g = \frac{1 + 2\cos \frac{2\pi l }{3} +2\tan \frac{\pi L}{3}\sin \frac{2\pi l}{3}}{3}\,.
\end{align}

\section{Lattice realization of edge theory of Levin-Gu state}
\label{sec:edge_lg_z4}
In this part, we will study 1D edge theory of the $\ZZ_2$ SPT phase by realizing it on 1D lattice.
Let us consider a 1D spin-$\frac{1}{2}$ chain, where local spins live at bond centers.
The anomalous $\ZZ_2$ symmetries is represented as the following non-onsite action\cite{spt_chen_mpo,spt_levin2012braiding}
\begin{equation}
    U_{\partial A}(g)=\prod_j \sigma^x_{j+\frac{1}{2}}\cdot\exp\left[ \frac{\ii\pi}{4}(1-\sigma^z_{j-\frac{1}{2}}\sigma^z_{j+\frac{1}{2}}) \right]
    \label{eq:lg_edge_sym}
\end{equation}

Under $U_{\partial A}(g)$, $\sigma^x_{j+\frac{1}{2}}\to\sigma^z_{j-\frac{1}{2}}\sigma^x_{j+\frac{1}{2}}\sigma^z_{j+\frac{3}{2}}$, and $\sigma^z_{j+\frac{1}{2}}\to -\sigma^z_{j+\frac{1}{2}}$.
Therefore, the transverse field Ising model symmetric under Eq.~\eqref{eq:lg_edge_sym} reads
\begin{equation}
    H_{\text{edge}}=\sum_{j=1}^L -h\left( \sigma^x_{j+\frac{1}{2}}-\sigma^z_{j-\frac{1}{2}}\sigma^x_{j+\frac{1}{2}}\sigma^z_{j+\frac{3}{2}} \right) - J\sigma^z_{j-\frac{1}{2}}\sigma^z_{j+\frac{1}{2}}
    \label{eq:lg_edge_ham}
\end{equation}

The edge theory can also be realized as the low-energy sector of certain Hamiltonian with on-site $\ZZ_4$ action by introducing extra qubits.
Specifically, we consider 1D chain with two qubits at each site $j$, where the $\ZZ_4$ symmetry is generated by 
\begin{equation}
    W_{\partial A}(g)\equiv\prod_j X_j^l X_j^r\cdot\exp\left[ \frac{\ii\pi}{4}(1-Z_j^l Z_j^r) \right]=\prod_j W_j(g)~,
    \label{eq:z4_gapless_spt_sym}
\end{equation}
and $W^2_{\partial A}(g)=\prod_j Z_j^lZ_j^r$.
When restricted to the sector where $Z_j^l=Z_{j+1}^r\equiv\sigma^z_{j+\frac{1}{2}}$, it is straightforward to check that the $\ZZ_4$ group reduce to $\ZZ_2$, and $W_{\partial A}(g)$ equals $U_{\partial A}(g)$ in Eq.~\eqref{eq:lg_edge_sym}.
Under $W_{\partial A}(g)$ action, $X_j^l\to Y_j^l Z_j^r$, $X_j^r\to Z_j^l Y_j^r$, $Y_j^l\to X_j^l Z_j^r$, $Y_j^r\to Z_j^l X_j^r$ and $Z_j^{l/r}\to-Z_{j}^{l/r}$.

We now consider the following Hamiltonian
\begin{equation}
    H_{\text{gSPT}}=\sum_{j=1}^L -h\left( X_j^r X_{j+1}^l+Z_j^lY_j^rY_{j+1}^lZ_{j+1}^r \right)-J Z_j^lZ_{j}^r
    +h_y\left( -Z^l_{j-1} X_j^r Y_{j+1}^l+Z^l_{j-1}Z_j^lY_j^rX_{j+1}^lZ_{j+1}^r \right)-\Delta Z_j^r Z_{j+1}^l
    \label{eq:z4_sym_lg_edge}
\end{equation}
It is straightforward to check that Eq.~\eqref{eq:z4_sym_lg_edge} invariant under $W_{\partial A}(g)$, and also commute with $Z_j^r Z_{j+1}^l$, $\forall j$.
For the parameter region $\Delta\gg J,h,h_y\ge0$, the low-energy sector is identified with $Z^r_jZ^l_{j+1}=1$.
Restricted to such low-energy sector and set $h_y=0$, $H_{\text{gSPT}}$ reduces to $H_{\text{edge}}$ in Eq.~\eqref{eq:lg_edge_ham}.

We further introduce time reversal operation $W_{\partial A}(\TT)\equiv\prod_j W_j(\mathcal{T})$ with $W_j(\mathcal{T}) = {\rm{diag}}(1,\,1,\,-1,\,1)$. 
It then commutes with $W(g)$:
\begin{equation}
    W_{j}(\mathcal{T})W_{j}(g) = [W_{j}(g)]^*W_{j}(\mathcal{T})
    \label{eq:gT_comm}
\end{equation}
Under the time reversal action, $X_j^l\to-X_j^lZ_j^r$, $X_j^r\to Z_j^l X_j^r$, $Y_j^l\to Y_j^lZ_j^r$, $Y_j^r\to -Z_j^lY_j^r$, and $Z_j^{l/r}\to Z_j^{l/r}$.
Hence, at the low-energy sector where $Z_j^r Z_{j+1}^l=1$, the $h$-term and the $J$-term preserve time reversal symmetry, while the $h_y$-term is odd under time reversal.

We also write down the system with boundary condition twisted by $g$-flux.
Therefore, terms across bond $L+\frac{1}{2}$ becomes
\begin{equation}
    -h\left( X_L^r Y_{1}^lZ_1^r-Z_L^lY_L^rX_{1}^l \right)-J Z_L^lZ_L^r 
    +h_y\left( -Z^l_{L-1}X_L^rX_1^lZ_1^r-Z^l_{L-1}Z_L^lY_L^rY_1^l -Z_L^lZ_1^lY_1^rX_2^lZ_2^r+Z_L^lX_1^rY_2^l \right)+\Delta Z_L^r Z_1^l
    \label{eq:lg_edge_twisted_term}
\end{equation}
while other terms remain the same as Eq.~\eqref{eq:z4_sym_lg_edge}.

Numerical results for the ground state degeneracy of Hamiltonian with various boundary conditions are listed in the following two tables.
\begin{center}
    \begin{tabular}{ |p{2cm}||p{2cm}|p{2cm}|p{2cm}|p{2cm}|  }
        \hline
        \multicolumn{5}{|c|}{GSD of untwisted $H_{\rm{gSPT}}$} \\
        \hline
        size & $J=0$, $h_y=0$ &$J\neq 0$, $h_y=0$&$J=0$, $h_y\neq 0$&$J\neq 0$, $h_y\neq 0$\\
        \hline
        $L=2n+1$   &1   &1   &1   &1\\
        $L=4n+2$   &2   &1   &2   &1\\
        $L=4n$     &1   &1   &1   &1\\
        \hline
    \end{tabular}
    \end{center}

\begin{center}
\begin{tabular}{ |p{2cm}||p{2cm}|p{2cm}|p{2cm}|p{2cm}|  }
    \hline
    \multicolumn{5}{|c|}{GSD of $g$-twisted $H_{\rm{gSPT}}$} \\
    \hline
    size & $J=0$, $h_y=0$ &$J\neq 0$, $h_y=0$&$J=0$, $h_y\neq 0$&$J\neq 0$, $h_y\neq 0$\\
    \hline
    $L=2n+1$   &2   &2   &1   &1\\
    $L=4n+2$   &2   &2   &2   &2\\
    $L=4n$     &4   &2   &2   &1\\
    \hline
\end{tabular}
\end{center}

Note that field theory prediction is only consistent with the case where $L=4n$ and $J=0$, but fails for other cases.

\section{The RDM for generic 2D SPT phases}
\label{sec:rdm_spt}
As mentioned in the main text, the RDM of a 2+1D SPT phase, when expressed in the entanglement space, belongs to a 1D intrinsic average SPT phase.
In this section, we formalize such correspondence in detail.

\subsection{Crossed modules}
Such correspondence is most naturally explained through \emph{a crossed module}.
Consider two groups $E$ and $N$, equipped with action of $E$ on $N$ (denoted as ${}^{e}\!n$), and a homomorphism $\varphi:N\to E$, satisfying 
\begin{equation}
    {}^{\varphi(n)}\!n'=nn'n^{-1}~,\quad
    \varphi\left( \,{}^{e}\!n \right)=e\varphi(n)e^{-1}~.
    \label{eq:crossed_module_cond}
\end{equation}
Here, $N$, together with $\varphi$ and the group action, form a crossed module over $E$.
From Eq.~\eqref{eq:crossed_module_cond}, $\operatorname{Im} \varphi$ is normal in $E$, allowing the definition $G\equiv\coker{\varphi}$.
Restricting to the case where $\ker{\varphi}=U(1)$ (central in $N$ by Eq.~\eqref{eq:crossed_module_cond}), we have the following exact sequence known as the crossed module extensions
\begin{align}
    1\rightarrow U(1)\rightarrow N\xrightarrow{\varphi} E\rightarrow G\rightarrow 1
    \label{eq:crossed_module_extension}
\end{align}
where $E$-action on $N$ induces a $G$-module structure on $U(1)$.
The above exact sequence can also be rewritten as two short exact sequences:
\begin{equation}
    \begin{aligned}
        &1\rightarrow U(1)\rightarrow N\xrightarrow{\phi} M\rightarrow 1\\
        &1\rightarrow M\xrightarrow{i} E\rightarrow G\rightarrow 1
    \end{aligned}
    \label{eq:decompose_crossed_module}
\end{equation}
where $M=\phi(N)$, $i:M\hookrightarrow E$ is an inclusion map, and $\varphi=i\circ\phi$.

It turns out~\cite{holt1979interpretation,huebschmann1980crossed,thomas2009third,brown2012cohomology,sptedge_else2014classifying} that such crossed module extensions are classified by $H^3(G,U(1))$, which also classifies 2D SPT phases protected by symmetry $G$. 
Therefore, given onsite unitary symmetry group $G$ and a third cohomology class $[\omega]\in H^3(G,U(1))$, there exists crossed modules characterized by $\omega$ 
\footnote{Generalizations to anti-unitary symmetries and spatial symmetries follow analogously}.

\subsection{Anomalous edge symmetry from crossed modules}
Let $\HH_{\partial A}$ denote Hilbert space associated with the 1D edge $\partial A$.
We now construct the anomalous $G$-symmetry action on $\HH_{\partial A}$ from cross module extension in Eq.~\eqref{eq:crossed_module_cond}, whose anomaly data is described by the cocycle $\omega$ characterizing Eq.~\eqref{eq:crossed_module_cond}.

More specifically, the second line in Eq.~\eqref{eq:decompose_crossed_module} describe a onsite projective representation of $G$ with coefficient in $M$, represented as $W_{\partial A}\equiv\prod_{j\in\partial A}W_j(g)$, where $W_j$ satisfies
\begin{equation}
    W_j(g_1)\cdot W_j(g_2)= D_j(g_1,g_2) W_j(g_1g_2)
    \label{eq:proj_rep}
\end{equation}
Here, $D_j$ is a representation of $M$ at site $j$, satisfying the two-cocycle condition due to associativity:
\begin{equation}
    D_j(g_1,g_2)\cdot D_j(g_1g_2,g_3)= \left( {}^{W_j(g_1)}\!D_j(g_2,g_3) \right)\cdot D_j(g_1,g_2g_3)
    \label{eq:two_cocycle_cond}
\end{equation}
To realize the first line in Eq.~\eqref{eq:decompose_crossed_module}, we introduce $\lambda_j^{l/r}$ as a representation of $N$, with the $U(1)$ phase factor as center.
The homomorphism $\phi$ in Eq.~\eqref{eq:decompose_crossed_module} is implemented by assuming the following decomposition 
\begin{equation}
    D_j=\lambda_j^l\cdot [\lambda_j^r]^\dg
    \label{eq:D_decomp}
\end{equation}
As such decomposition hosts $U(1)$ phase ambiguity with $\lambda_j^{l/r}\to\ee^{\ii\theta}\lambda_j^{l/r}$, lifting Eq.~\ref{eq:two_cocycle_cond} to $\lambda^{l/r}$ introduces a 3-cocycle $\omega$ characterizing the crossed module:
\begin{equation}
    \lambda_j^{l/r}(g_1,g_2)\cdot \lambda_j^{l/r}(g_1g_2,g_3)
    =\omega(g_1,g_2,g_3)\cdot (W_j(g_1)\circ \lambda_j^{l/r}(g_2,g_3)) \cdot \lambda_{j}^{l/r}(g_1,g_2g_3)
    \label{eq:lift_two_cocycle}
\end{equation}
The three-cocycle condition can be obtained by fusing $\lambda^l(g_1,g_2)\lambda^l(g_1g_2,g_3)\lambda^l(g_1g_2g_3,g_4)$ in two different ways.

By imposing local constraint $[\lambda_j^r]^\dg\lambda_j^l=1~\forall j$, one obtains a subspace of $\HH_{\partial A}$.
In this subspace $\prod_j D_j=1$ due to Eq.~\eqref{eq:D_decomp}, and thus $W_{\partial A}(g)$ becomes a linear representation of $G$ according to Eq.~\eqref{eq:proj_rep}. 
One can further demonstrate that such linear representation manifests as the MPO that encodes the anomaly data $\omega$.

\subsection{RDM in the entanglement space and tensor constructions}
We now consider an SPT wavefunction $\ket{\psi}$ expressed by PEPS, bipartition the system along $\partial A$, and identify $\HH_{\partial A}$ as the virtual legs along $\partial A$.
The global symmetry $g\in G$ is implemented as Eq.~\eqref{eq:schmidt_tensor_sym_sm}, while $D_{\partial A}$, or more precisely $[\lambda_j^r]^\dg\lambda_{j+1}^l$ follows the implementation in Eq.~\eqref{eq:schmidt_tensor_igg_sm}.
Therefore, $W_{\partial A}(g)$ acts as a weak symmetry, while $D_{\partial A}(g_1,g_2)$ is a strong symmetry on $\rho_{\partial A}$, and $\lambda$'s impose the local constraint:
\begin{align}
    \rho_{\partial A}=[\lambda_j^r]^\dg\lambda_{j+1}^l \cdot \rho_{\partial A}
    =\rho_{\partial A} \cdot [\lambda_j^r]^\dg\lambda_{j+1}^l 
    \label{eq:rho_local_constraint}
\end{align}
The above equations, together with the short-range correlated condition, make $\rho_{\partial A}$ an intrinsic average SPT phase protected by weak symmetry $E$ and its strong symmetry subgroup $N$\cite{mix_ma2023intrinsic}. 
It is characterized by the following string order parameter:
\begin{align}
    \lim_{\abs{l}\to\infty}\Tr\left[ \rho_{\partial A}\cdot \left([\lambda_1^r(g_1,g_2)]^\dg\cdot \left( \prod_{j=2}^{l}D_{j}(g_1,g_2) \right)\cdot \lambda_{l+1}^l(g_1,g_2)\right) \right]=1
    \label{}
\end{align}

We now outline the tensor construction for variational wavefunctions of the SPT phase characterized by three cocycle $\omega$ (see Ref.~\cite{spt_jiang2016} for details).
While $W(g)$ and $\lambda^{l/r}$ are interpreted as symmetry constraints on $\HH_{\partial A}$, they can also be implemented as symmetry conditions on local tensors in PEPS.
Namely, by identifying $A$ as a single site, and $\HH_{\partial A}$ as virtual legs of a local tensor, we impose the global symmetry Eq.~\eqref{eq:schmidt_tensor_sym_sm} and virtual leg constraint Eq.~\eqref{eq:schmidt_tensor_igg_sm} on the local tensor.
Here, $W(g)$ and $\lambda^{l/r}$ satisfy algebraic relations listed in Eq.~\eqref{eq:proj_rep}, Eq.~\eqref{eq:D_decomp} and Eq.~\eqref{eq:lift_two_cocycle}.
The solution space for the local tensor then give variational PEPS ansatz for the SPT phase.

\subsection{Example: the $\ZZ_2$ SPT phase}
We now fit the $\ZZ_2$ SPT phase~(analyzed in the main text) into the present framework based on crossed modules.
For this case, $N=\ZZ_2\times U(1)$, generated by $\lambda^{l/r}\ee^{\ii\theta}$; $M=\ZZ_2$, generated by operator $D$; $E=\ZZ_4$, generated by $W(g)$.
The action of $E$ on $N$ is determined by the algebraic relation $W(g)\circ \lambda^{l/r}=-\lambda^{l/r}$.

This construction admits a natural anyon condensation interpretation (also works for more general cases beyond $G=\ZZ_2$~\cite{spt_jiang2016,ellison2022pauli}). 
The starting point is a symmetry-enriched topological~(SET) phase with 
\begin{enumerate}
    \item $\ZZ_2$ toric code topological order, where $D\in M$ creates $\ZZ_2$ flux line ($m$-anyons at endpoints)
    \item $G=\ZZ_2$ global symmetry 
    \item $E=\ZZ_4$ realized by $[W(g)]^2=D$, implying that fusion of two $g$-defect gives $m$-anyon

Equivalently it implies that $e$-anyon carries half charge of symmetry $G$ in standard SET language, as braiding $g$-defect on $e$-anyon twice yields $(-1)$ Berry phase.
\end{enumerate}
The map $\phi: N\to M$ encodes the condensation of bound states of $m$-anyons and $\lambda^{l/r}$, while the action of $E$ on $N$ implies that $\lambda^{l/r}$ is charged under $G$. 
This condensation drives the SET phase to the desired $\ZZ_2$ SPT phase.

In the following, we list the dictionary between crossed modules, anyon condensation interpretation, PEPS construction, and the intrinsic average SPT for the $\ZZ_2$ case.

\begin{table}[h]
\centering
\caption{}
\label{tab:z2_spt_dictionary}
\begin{tabular}{|c|c|c|c|}
\hline
\textbf{Crossed module} & \textbf{Anyon condensation} & \textbf{Symmetries in PEPS} & \textbf{Intrinsic average SPT phase}\\
\hline
$G = \ZZ_2=\{1,g\}$ & \makecell{$\ZZ_2$ global symmetry} & \makecell{Physical $\ZZ_2$ symmetry \\ generated by $U(g)$} & \makecell{Quotient of weak and \\ strong symmetry}  \\
\hline
$M = \ZZ_2$ & $\ZZ_2$ toric code topological order & Virtual $\ZZ_2$ symmetry generated by $D$ & Strong $\ZZ_2$ symmetry  \\
\hline
$N = \mathbb{Z}_2 \times U(1)$ & Local excitation $\lambda\ee^{\ii\theta}$ with $\lambda^2=\hat{1}$ & \makecell{Virtual $\ZZ_2$ symmetry~($\lambda^{l/r}$) \\ within a plaquette} & Local projection $\frac{1}{2}(1+\lambda_{j}^r\lambda_{j+1}^l)$ \\
\hline
$E = \mathbb{Z}_4$ & $e$ carries half $\ZZ_2$ symmetry charge& \makecell{Virtual symmetry $W(g)$ \\ with $[W(g)]^2=D$} & Weak $\ZZ_4$ symmetry \\
\hline
Action of $E$ on $N$ & $\lambda$ carries unit symmetry charge & $W(g)\lambda^{l/r} = -\lambda^{l/r}W(g)$ &  $\lambda$ carries double-$\ZZ_4$ charge\\
\hline
$\phi: N \to M$ & \makecell{Condensing bound state \\ of $m$ and $\lambda$} & $D=\lambda^l\cdot\lambda^r$  & \makecell{Bind strong $\ZZ_2$ symmetry flux \\ with $\lambda$} \\
\hline
\end{tabular}
\end{table}

\section{Details in MPDO-based proof}
\label{sec:mpdo_proof}
Let us analyze symmetries of $T(g)$ in detail.
The weak symmetry condition $\rho_{\partial A}=W_{\partial A}(g)\cdot \rho_{\partial A}\cdot W_{\partial A}^\dg$ for the MPDO $\rho_{\partial A}$ imposes constraint on local tensor $M$ 
\begin{align*}
    \begin{tikzpicture}[scale=1,baseline={([yshift=-2pt]current bounding box.center)}]
        \draw (8pt,-15pt) -- (8pt,15pt);
        \draw (-7pt,0pt) -- (23pt,0pt);
        \filldraw[thick,fill=white] (0pt,-8pt) rectangle (16pt,8pt);
        \node at (8pt,0pt) {$M$};
    \end{tikzpicture}
    =
    \begin{tikzpicture}[scale=1,baseline={([yshift=-2pt]current bounding box.center)}]
        \draw (8pt,-20pt) -- (8pt,20pt);
        \draw (-12pt,0pt) -- (28pt,0pt);
        \filldraw[thick,fill=white] (0pt,-8pt) rectangle (16pt,8pt);
        \node at (8pt,0pt) {$M$};
        \filldraw [red,fill=red] (8pt,15pt) circle (1pt) node[left] {\scriptsize $W(g)$};
        \filldraw [red,fill=red] (8pt,-15pt) circle (1pt) node[left] {\scriptsize $W^\dagger(g)$};
        \filldraw [red,fill=red] (-9pt,0pt) circle (1pt) node[left] {\scriptsize $V(g)$};
        \filldraw [red,fill=red] (25pt,0pt) circle (1pt) node[right] {\scriptsize $V^\dg(g)$};
    \end{tikzpicture};
\end{align*}
where $V(g)$ is some gauge transformation on the internal leg of the MPDO.
It further provides symmetries of $T(g)$
\begin{align}
    T(g)=V(g)\cdot T(g)\cdot V^\dg(g)
    \label{eq:twisted_tm_z2_sym_app}
\end{align}
When treating  
\begin{equation}
    \rho_{\partial A}
    =\lambda_{j}^r\lambda_{j+1}^l\cdot \rho_{\partial A}
    =\rho_{\partial A}\cdot \lambda_{j}^r\lambda_{j+1}^l
    \label{eq:rho_bdry_plq_igg}
\end{equation}
as symmetries, we obtain constraint for the local tensor $M$:
\begin{align*}
    \begin{tikzpicture}[scale=0.75,baseline={([yshift=-2pt]current bounding box.center)}]
        \draw (8pt,-20pt) -- (8pt,20pt);
        \draw (-12pt,0pt) -- (28pt,0pt);
        \filldraw[thick,fill=white] (0pt,-8pt) rectangle (16pt,8pt);
        \node at (8pt,0pt) {$M$};
    \end{tikzpicture}
    ={}&
    \begin{tikzpicture}[scale=0.75,baseline={([yshift=-2pt]current bounding box.center)}]
        \draw (8pt,-20pt) -- (8pt,20pt);
        \draw (-12pt,0pt) -- (28pt,0pt);
        \filldraw[thick,fill=white] (0pt,-8pt) rectangle (16pt,8pt);
        \node at (8pt,0pt) {$M$};
        \filldraw [red,fill=red] (8pt,15pt) circle (1pt) node[left] {\scriptsize $\lambda^l$};
        \filldraw [red,fill=red] (-9pt,0pt) circle (1pt) node[left] {\scriptsize $\lambda^u$};
    \end{tikzpicture} 
    =
    \begin{tikzpicture}[scale=0.75,baseline={([yshift=-2pt]current bounding box.center)}]
        \draw (8pt,-20pt) -- (8pt,20pt);
        \draw (-12pt,0pt) -- (28pt,0pt);
        \filldraw[thick,fill=white] (0pt,-8pt) rectangle (16pt,8pt);
        \node at (8pt,0pt) {$M$};
        \filldraw [red,fill=red] (8pt,15pt) circle (1pt) node[left] {\scriptsize $\lambda^r$};
        \filldraw [red,fill=red] (25pt,0pt) circle (1pt) node[right] {\scriptsize $(\lambda^{u})^\dg$};
    \end{tikzpicture} \\
    ={}&
    \begin{tikzpicture}[scale=0.75,baseline={([yshift=-2pt]current bounding box.center)}]
        \draw (8pt,-20pt) -- (8pt,20pt);
        \draw (-12pt,0pt) -- (28pt,0pt);
        \filldraw[thick,fill=white] (0pt,-8pt) rectangle (16pt,8pt);
        \node at (8pt,0pt) {$M$};
        \filldraw [red,fill=red] (8pt,-15pt) circle (1pt) node[left] {\scriptsize $(\lambda^l)^\dg$};
        \filldraw [red,fill=red] (-9pt,0pt) circle (1pt) node[left] {\scriptsize $\lambda^d$};
    \end{tikzpicture} 
    =
    \begin{tikzpicture}[scale=0.75,baseline={([yshift=-2pt]current bounding box.center)}]
        \draw (8pt,-20pt) -- (8pt,20pt);
        \draw (-12pt,0pt) -- (28pt,0pt);
        \filldraw[thick,fill=white] (0pt,-8pt) rectangle (16pt,8pt);
        \node at (8pt,0pt) {$M$};
        \filldraw [red,fill=red] (8pt,-15pt) circle (1pt) node[left] {\scriptsize $(\lambda^r)^\dg$};
        \filldraw [red,fill=red] (25pt,0pt) circle (1pt) node[right] {\scriptsize $(\lambda^{d})^\dg$};
    \end{tikzpicture} 
\end{align*}
As $W(g)$ and $\lambda^{l/r}$ anti-commute, for $T(g)$, we have
\begin{align}
    T(g)=-\lambda^{u}\lambda^d\cdot T(g)
    =-T(g)\cdot \lambda^{u}\lambda^d
    \label{eq:twisted_tm_plq_igg}
\end{align}

We next analyze the algebraic relations between symmetry actions on the internal legs of $\rho_{\partial A}$.
The simplest case is that these algebraic relations mirror those on the physical legs of the MPDO, i.e.
\begin{equation}
    \begin{aligned}
        &[V(g)]^2=\lambda^u\lambda^d~,\quad
        [\lambda^{u/d}]^2=1\\
        &V(g)\cdot\lambda^{u/d}=-\lambda^{u/d}\cdot V(g)\\
    \end{aligned}
    \label{eq:rho_vleg_tensor_eq_sm}
\end{equation}
One may worry that symmetries on internal legs form projective representations, similar to the 1D SPT phase. 
However, nontrivial projective representation will lead to degenerate eigenvalues of $T$, violating our assumption.
A more subtle case is that such internal leg symmetries satisfy Eq.~\eqref{eq:rho_vleg_tensor_eq_sm} up to a operator.
That means there exist symmetry acting solely on internal legs, which gives non-decaying correlator measured in $\rho^2$.
Such case resembles strong-to-weak symmetry breaking in mixed state\cite{mix_lessa2024strong}, which is not consistent for the RDM of SRE phase.
Therefore, we will use Eq.~\eqref{eq:rho_vleg_tensor_eq_sm} for the following derivations.

From Eq.~\eqref{eq:twisted_tm_z2_sym_app}, \eqref{eq:twisted_tm_plq_igg}, and \eqref{eq:rho_vleg_tensor_eq_sm}, given an eigenstate $\vket{\psi}$ of $T(g)$ with non-zero eigenvalue $\alpha$, we have
\begin{align}
    [V(g)]^2\vket{\psi}=\alpha^{-1}\cdot \lambda^u\lambda^d\cdot T(g)\vket{\psi}=-\vket{\psi}
    \label{}
\end{align}
where we use Eq.~\eqref{eq:twisted_tm_plq_igg} to get the last equal sign.
Therefore, 
\begin{align}
    V(g)\vket{\psi}=\pm\ii\vket{\psi}
    \label{eq:twisted_tm_eigen_z2_qn_sm}
\end{align}

Now, we introduce the Hermiticity condition of $\rho_{\partial A}$.
Here, Hermitian conjugate can be viewed as an anti-unitary operator on ket and bra space of $\rho_{\partial A}$, which exchanges ket and bra states.
Therefore, $\rho_{\partial A}=\rho_{\partial A}^\dg$ imposes an anti-unitary symmetry condition on $M$ with $V(\mathcal{H})$ acting on internal legs:
\begin{align*}
    \begin{tikzpicture}[scale=1,baseline={([yshift=-2pt]current bounding box.center)}]
        \draw (8pt,-15pt) -- (8pt,15pt);
        \draw (-12pt,0pt) -- (28pt,0pt);
        \filldraw[thick,fill=white] (0pt,-8pt) rectangle (16pt,8pt);
        \node at (8pt,0pt) {$M$};
        \node[above] at (8pt,15pt) {\scriptsize $\alpha$};
        \node[below] at (8pt,-15pt) {\scriptsize $\beta$};
    \end{tikzpicture}
    =
    \begin{tikzpicture}[scale=1,baseline={([yshift=-2pt]current bounding box.center)}]
        \begin{scope}[]
            \draw (8pt,-15pt) -- (8pt,15pt);
            \draw (-12pt,0pt) -- (28pt,0pt);
            \filldraw[thick,fill=white] (0pt,-8pt) rectangle (16pt,8pt);
            \node at (8pt,0pt) {$M^*$};
            \node[above] at (8pt,15pt) {\scriptsize $\beta$};
            \node[below] at (8pt,-15pt) {\scriptsize $\alpha$};
            \filldraw [red,fill=red] (-9pt,0pt) circle (1pt) node[left] {\scriptsize $V(\mathcal{H})$};
            \filldraw [red,fill=red] (25pt,0pt) circle (1pt) node[right] {\scriptsize $V^\dg(\mathcal{H})$};
        \end{scope}
    \end{tikzpicture}
\end{align*}
Consequently, $T(g)$ hosts the following anti-unitary symmetry:
\begin{equation}
    \begin{aligned}
        T(g)&=V(\mathcal{H})\cdot (\Tr[W^\dg(g)\cdot M])^*\cdot V^\dg(\mathcal{H})\\
        &=V(\mathcal{H})\cdot (\Tr[D\cdot W(g)\cdot M])^*\cdot V^\dg(\mathcal{H})\\
        &=V(\mathcal{H})\cdot(\lambda^u)^*\cdot [T(g)]^* \cdot(\lambda^u)^\tp\cdot V^\dg(\mathcal{H})
    \end{aligned}
    \label{}
\end{equation}
Note that under Hermitian conjugate, action of weak symmetry is invariant, while strong symmetry on ket space transforms to that on bra space.
Symmetry relation on internal legs follows those on physical legs, giving
\begin{equation}
    \begin{aligned}
        &V(\cH)\cdot V^*(\cH)=1~,\\
        &V(g)\cdot V(\mathcal{H})=V(\mathcal{H})\cdot V^*(g)~,\\
        &\lambda^u\cdot V(\mathcal{H})=V(\mathcal{H})\cdot (\lambda^d)^*
    \end{aligned}
    \label{eq:rho_vleg_hermitian_sym_relation}
\end{equation}
Using the last line of the above equation, together with Eq.~\eqref{eq:twisted_tm_plq_igg}, we obtain 
\begin{align}
    \left[ V(\mathcal{H})\KK (\lambda^u) \right]^2\cdot T(g)=-T(g)
    \label{}
\end{align}
According to Kramers theorem, eigenstates of $T(g)$ are at least doubly degenerate (in terms of module): for an eigenstates $\vket{\psi}$ of $T(g)$ with eigenvalue $\alpha$, $V(\cH)\mathcal{K}\lambda^u\vket{\psi}$ is another eigenstate with eigenvalue $\alpha^*$.
The doubly degenerate dominant eigenvalues of $T(g)$ would give anomalous contribution in $-\ln \lrangle{U_A(g)} = -\ln \Tr [(T(g))^{L_{\partial A}}]$ in the limit $L_{\partial A}\to \infty$.

Besides, from Eq.~\eqref{eq:rho_vleg_tensor_eq_sm} and \eqref{eq:rho_vleg_hermitian_sym_relation}, we have
\begin{align}
    V(g)\cdot V(\cH)\KK\lambda^u\vket{\psi}=-V(\cH)\KK\lambda^u V(g)\vket{\psi}
    \label{eq:twisted_eigen_hermitian_doublet_z2_qn}
\end{align}
Furthermore, due to Eq.~\eqref{eq:twisted_tm_eigen_z2_qn_sm}, we conclude that the doublet eigenstates $\vket{\psi}$ and $V(\cH)\KK\lambda^u\vket{\psi}$ share the same quantum number under $V(g)$.


If the system further hosts time reversal symmetry $\rho_{\partial A}= W_{\partial A}(\TT)\cdot \rho^*_{\partial A}\cdot W^\dg_{\partial A}(\TT)$, local tensor should satisfy the following constraint:
\begin{align*}
    \begin{tikzpicture}[scale=1,baseline={([yshift=-2pt]current bounding box.center)}]
        \draw (8pt,-15pt) -- (8pt,15pt);
        \draw (-12pt,0pt) -- (28pt,0pt);
        \filldraw[thick,fill=white] (0pt,-8pt) rectangle (16pt,8pt);
        \node at (8pt,0pt) {$M$};
    \end{tikzpicture}
    =
    \begin{tikzpicture}[scale=1,baseline={([yshift=-2pt]current bounding box.center)}]
        \draw (8pt,-20pt) -- (8pt,20pt);
        \draw (-12pt,0pt) -- (28pt,0pt);
        \filldraw[thick,fill=white] (0pt,-8pt) rectangle (16pt,8pt);
        \node at (8pt,0pt) {$M^*$};
        \filldraw [red,fill=red] (8pt,15pt) circle (1pt) node[left] {\scriptsize $W(\TT)$};
        \filldraw [red,fill=red] (8pt,-15pt) circle (1pt) node[left] {\scriptsize $W^\dg(\TT)$};
        \filldraw [red,fill=red] (-9pt,0pt) circle (1pt) node[left] {\scriptsize $V(\mathcal{T})$};
        \filldraw [red,fill=red] (25pt,0pt) circle (1pt) node[right] {\scriptsize $V^\dg(\TT)$};
    \end{tikzpicture} 
\end{align*}
which leads to 
\begin{align}
    T(g)=V(\TT)\cdot [T(g)]^*\cdot V^\dg(\TT)
    \label{}
\end{align}
Note that group relation between $V(\TT)\KK$ and other symmetry operations on internal legs follows those on physical legs of $\rho_{\partial A}$.
In particular,
\begin{align}
    V(\TT)V^*(g)=V(g)V(\TT)
    \label{}
\end{align}
Following the analysis presented subsequent to Eq.~\eqref{eq:twisted_eigen_hermitian_doublet_z2_qn}, if $\vket{\psi}$ is an eigenstate of $T(g)$ with quantum number $\pm\ii$ under $V(g)$, then $V(\TT)\KK\vket{\psi}$ would be a different eigenstate, possessing opposite quantum number under $V(g)$.

To summarize, let $\vket{\psi}$ to be the dominant eigenstate of $T(g)$ with eigenvalue $\ee^{-\alpha+\ii\theta}$ and $V(g)\vket{\psi}=\ii\vket{\ii}$.
The four states $\left\{ \vket{\psi_1}\equiv\vket{\psi},~\vket{\psi_2}\equiv V(\HH)\KK\vket{\psi},~\vket{\psi_3}\equiv V(\TT)\KK\vket{\psi},~\vket{\psi_4}\equiv V(\HH)V(\TT)\vket{\psi} \right\}$ are all eigenstates of $T(g)$ with eigenvalues $\left\{ \ee^{-\alpha+\ii\theta},\ee^{-\alpha-\ii\theta},\ee^{-\alpha-\ii\theta},\ee^{-\alpha+\ii\theta} \right\}$ and $g$-quantum number $\left\{ \ii,-\ii,\ii,-\ii \right\}$, respectively.

We are now ready to calculate correlators of charged operator on symmetry twisted density operator $\rho^g_{\partial A}$.
For large $L_{\partial A}$, 
\begin{equation}
    T(g)\sim \sum_{a=1}^4 \xi_a\cdot\vket{\psi_a}\vbra{\psi_a}
    =\ee^{-\alpha+\ii\theta}\cdot \left(\vket{\psi_1}\vbra{\psi_1}+\vket{\psi_4}\vbra{\psi_4}\right) + \ee^{-\alpha-\ii\theta}\cdot \left(\vket{\psi_2}\vbra{\psi_2}+\vket{\psi_3}\vbra{\psi_3} \right)
\end{equation}
where $\xi_a$ labels eigenvalue for $\vket{\psi_a}$.
Therefore,
\begin{equation}
    \Tr(\rho_{\partial A}^g)\sim\sum_{a=1}^4 \xi_a^{L_{\partial A}}
    =\exp(-\alpha L_{\partial A})\cdot 4\cos(L_{\partial A}\theta)
\end{equation}
\begin{equation}
    \Tr(O_j O_k\rho_{\partial A}^g)\sim \sum_{a,b=1}^4 \xi_a^{L-l-1}\cdot\xi_b^{l-1}\cdot \vbra{\psi_a}T(g,O)\vket{\psi_b}\cdot \vbra{\psi_b}T(g,O)\vket{\psi_a}
\end{equation}
where $l=\abs{k-j}$, and $W(g)\cdot O=-O\cdot W(g)$.
Here,
\begin{equation}
    T(g,O)=
    \begin{tikzpicture}[scale=1,baseline={([yshift=-2pt]current bounding box.center)}]
        \draw (8pt,-20pt) -- (8pt,30pt);
        \draw (-8pt,0pt) -- (24pt,0pt);
        \draw (8pt,30pt) arc (180:0:2.5pt);
        \draw (8pt,-20pt) arc (180:360:2.5pt);
        \draw [densely dotted] (13pt,30pt) -- (13pt,8pt);
        \draw [densely dotted] (13pt,-20pt) -- (13pt,-8pt);
        \filldraw[thick,fill=white] (0pt,-8pt) rectangle (16pt,8pt);
        \node at (8pt,0pt) {$M$};
        \filldraw [red,fill=red] (8pt,15pt) circle (1pt) node[left] {\scriptsize $W(g)$};
        \filldraw [red,fill=red] (8pt,25pt) circle (1pt) node[left] {\scriptsize $O$};
    \end{tikzpicture}
\end{equation}
We note that $\vbra{\psi_a}T(g,O)\vket{\psi_b}\neq0$ only when $\vket{\psi_a}$ and $\vket{\psi_b}$ have opposite $g$-quantum number.
Therefore, for the trivial symmetric phase or even the $\ZZ_2$ SPT phase without time reversal symmetry, we expect $\Tr(O_j O_k\rho_{\partial A}^g)/\Tr(\rho_{\partial A})\to 0$ when $l\to \infty$.
In contrast, forthe time-reversal symmetric $\ZZ_2$ SPT phase,
\begin{equation}
    \lrangle{O_j O_k}^g_{A}=\frac{\Tr(O_j O_k\rho_{\partial A}^g)}{\Tr(\rho_{\partial A}^g)}
    \sim C(L_{\partial A}\theta, l\theta)
\end{equation}
where $C(L_{\partial A}\theta, l\theta)$ retains non-vanishing $O(1)$ oscillatory amplitude in the thermodynamic limit.

\bibliography{lg.bib}